# Universal mechanism of luminescence enhancement in doped perovskite nanocrystals from symmetry analysis

Ghada H. Ahmed[1], Yun Liu[2], Ivona Bravić[2], Xejay Ng[2], Ina Heckelmann[2], Pournima Narayanan[1], Martin. S. Fernández[1], Bartomeu Monserrat[2,3], Daniel N. Congreve[1], Sascha Feldmann[2,4*]

[1]Department of Electrical Engineering, Stanford University, Stanford, CA 94305, USA
[2]Cavendish Laboratory, University of Cambridge, Cambridge, CB30HE, UK
[3]Department of Materials Science and Metallurgy, University of Cambridge, Cambridge, CB30FS, UK
[4]Rowland Institute, Harvard University, Cambridge, MA 02142, USA
*Email: sfeldmann@rowland.harvard.edu

**Abstract**

Metal-halide perovskite nanocrystals have demonstrated excellent optoelectronic properties for light-emitting applications. Isovalent doping with various metals ($M^{2+}$) can be used to tailor and enhance their light emission. Although crucial to maximize performance, an understanding of the universal working mechanism for such doping is still missing. Here, we directly compare the optical properties of nanocrystals containing the most commonly employed dopants, fabricated under identical synthesis conditions. We show for the first time unambiguously and supported by first principles calculations and molecular orbital theory that element-unspecific symmetry-breaking rather than element-specific electronic effects dominate these properties under device-relevant conditions. The impact of most dopants on the perovskite electronic structure is predominantly based on local lattice periodicity breaking and resulting charge carrier localization, leading to enhanced radiative recombination, while dopant-specific hybridization effects play a secondary role. Our results suggest specific guidelines for selecting a dopant to maximize the performance of perovskite emitters in the desired optoelectronic devices.

Nanocrystalline materials based on metal-halide perovskites like $CsPbX_3$ (X = Cl, Br, I) have been shown to act as efficient solution-processable emitters for solid-state lighting and displays[1,2], and are promising for emerging quantum light applications based on single-photon[3] or spin-polarized emission[4,5]. Atomic doping of these materials, mostly based on substitution of the $Pb^{2+}$ ion for isovalent metal cations, has helped to further improve spectral tunability and photoluminescence quantum efficiency (PLQE), particularly in the blue spectrum where efficient light-emitting diodes (LEDs) remain a pressing goal[6–9]. Recently, it was shown that for the case of manganese ($Mn^{2+}$) doping the observed efficiency gains are the result of not only a reduction of non-radiative charge trapping but also of dopant-induced carrier localization, resulting in enhanced radiative recombination rates[10]. However, a mechanism of how doping influences the optical properties in perovskite nanocrystals with other dopants is still missing, and with it any generalizable understanding which captures all observed effects. Yet, it is this overarching mechanism that will be essential to maximize the performance of doped perovskite nanocrystals and guide the informed choice of doping element for device applications on a case-by-case basis.

For the first time, we directly measure, model, and compare the optical properties of the most effective doped perovskite nanocrystal compositions currently employed ($Mn^{2+}$, $Ni^{2+}$, $Zn^{2+}$, for $Cl^-$ and $Br^-$ halide environments, and further all alkaline earth metals), fabricated under identical synthesis conditions. We can thus largely exclude any deviations between chemical compositions and experimental measurement artefacts, which are often inevitable when comparing results from different labs. We find that the observed properties of all doped systems investigated can be well understood *via* a delicate interplay of largely dopant-independent structural effects resulting from lattice periodicity breaking, which induce bandgap widening and radiative rate increase on the one hand, and dopant-dependent chemical and electronic effects arising from orbital hybridization, on the other hand. Our mechanistic insights provide direct guidelines for the design of the most efficient emitters for a given device application by exploiting the synthetically chemical space.

**Results**

A series of the most commonly employed doped perovskite nanocrystals based on the transition metal ions $Mn^{2+}$, $Ni^{2+}$, and $Zn^{2+}$ isovalent to $Pb^{2+}$ was synthesized (see Supporting Information for details) and their steady-state absorbance and photoluminescence were characterized (Fig. 1a). Further below we show that the same effects can also explain doping with, for example, alkaline-earth metals (see Supporting Figs. S7-10 for a detailed investigation on those). We study here doping concentrations of 1-2 atomic-%, since these have demonstrated the highest PLQE (Supporting Fig. S4), while higher concentrations overly distort the perovskite host structure and induce additional non-radiative trap channels. Importantly, doping has been reported to induce a blue-shift of the optical bandgap in these materials[7]. To decouple the dopant-induced changes to the radiative rate due to lattice periodicity breaking from those purely originating from a rate increase induced by the blue-shifted bandgap, we synthetically calibrated all doped materials in such a way that they display very similar optical bandgaps compared to the undoped material.

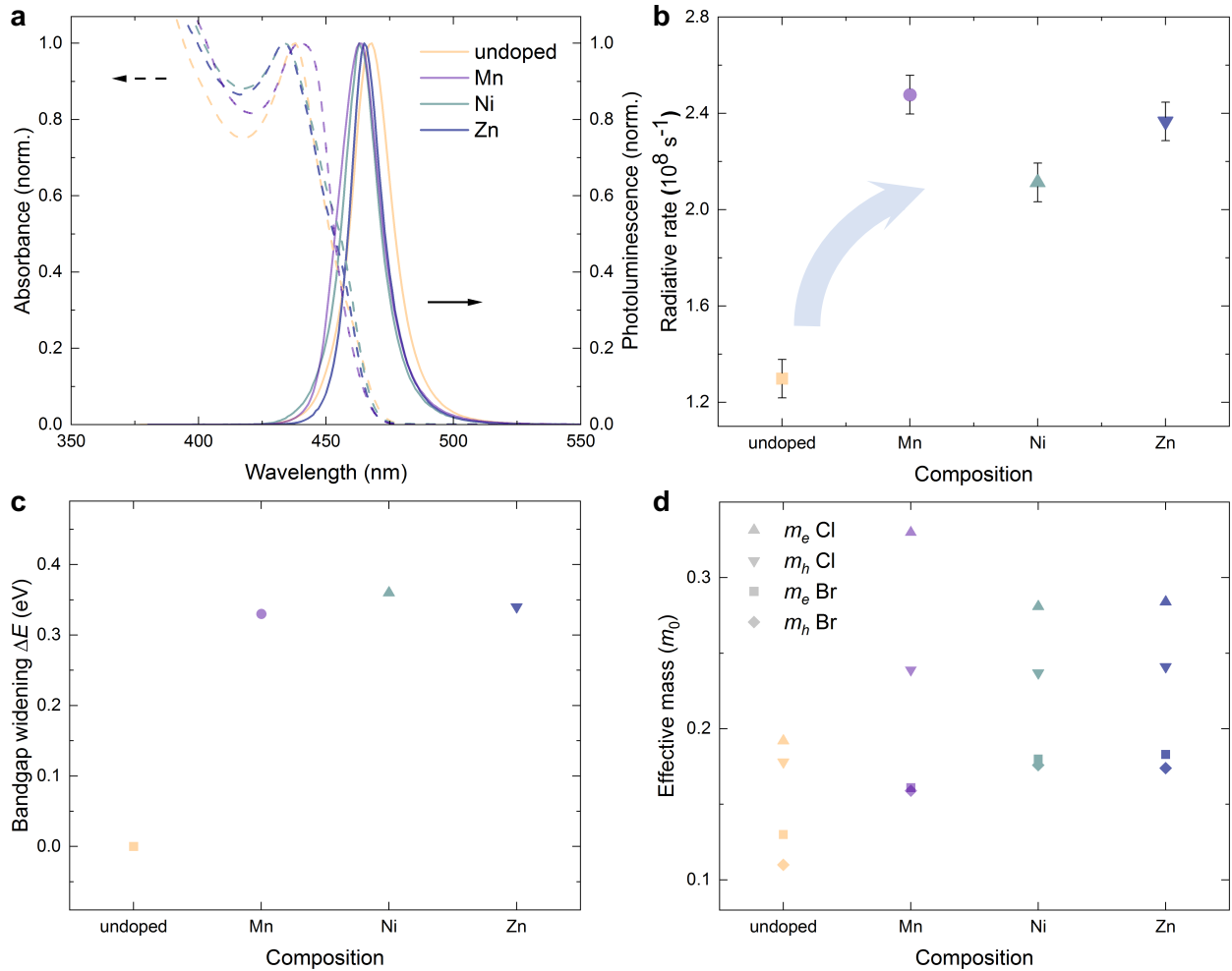

**Fig. 1 | Doping-enhanced optoelectronic properties of perovskite nanocrystals. a**, Absorbance (dashed) and photoluminescence (PL) spectra (continuous lines) of undoped and doped $CsPbCl_xBr_{3-x}$ NCs, chemically calibrated to have similar emission wavelength. Doping concentration is 1.5±0.5 atomic-%. **b**, Radiative recombination rates extracted from PLQE and time-resolved PL data (see SI for details). The $Mn^{2+}$ rate is extrapolated to correspond to the same concentration as for the other dopants. All samples were photoexcited at 405 nm. **c**, Calculated dopant-induced electronic bandgap widening without synthetic tuning through the halide composition (see Methods Section in SI for details). **d**, Calculated effective masses of different dopant-halide systems, increasing upon doping and reproducing all measured trends remarkably well.

For this, we first dope the perovskite with the respective dopant ion, which consequently blue-shifted the observed emission. In order to spectrally calibrate all samples to have a very similar emission wavelength, we added a few mL of lead bromide stock solution to each composition until the emission maximum was red-shifted back to 470 nm. We chose an emission of approximately 470 nm as this corresponds to an essential color coordinate for blue LEDs in display applications, as the y-value 0.08 on the Commission International de l'Eclairage1931 (CIE 1931) chromaticity diagram forms the primary blue standard according to the National Television System Committee[11]. The doping concentration of the final resulting NCs was determined with inductively-coupled plasma mass spectrometry to be 1.5±0.5 atomic-%, respectively (see Supporting Information). All studied nanocrystals were measured to be dopant-independent cubic shaped with an average size of 10±2 nm (see Supporting Fig. S1 for scanning electron microscopy images), implying the charge dynamics are situated within the typical weak quantum confinement regime[12], as is also confirmed by a linear PL dependence on excitation fluence (confirming excitonic recombination, see Supporting Fig. S2). Thus, no changes to the dielectric environment by NC shape or size could obscure the conclusions drawn from doping-induced changes to the optoelectronic properties, which we consequently investigate.

By measuring both the PLQE and the time-resolved PL decay of the nanocrystals we readily quantify and compare the radiative recombination rate for each composition (Fig. 1b, see Supporting Information for details on the calculations and Figs. S3-4 for underlying data). Importantly, for all doped compositions we find a – now bandgap-unrelated, as spectrally calibrated – substantial radiative rate increase of 63-92% compared to the undoped NCs. This generalizes the initial observation made for Mn-doping before[10] where we showed that this rate increase directly relates to an increase in the oscillator strength of the electronic transition (see Supporting Fig. S5 for oscillator strength values). We stress that these findings are distinctly different from the limited observation of increasing PLQEs upon doping, which is mostly assigned to trap-passivation in the literature, *e.g.* by filling halide vacancies[7]. Such reduction of non-radiative rates can be achieved by external changes to the semiconductor, for example by surface passivation or stability approaches and can dramatically improve PLQEs[13,14]. It leaves, however, the highest achievable light emission for a given semiconductor untouched, as the intrinsic radiative rate remains unchanged in this case, and comes with drawbacks for devices, *e.g.* through unfavorable work function shifts[15], or reduced carrier mobilities if long insulating ligands are employed. Moreover, PLQEs vary largely between different synthesis conditions for seemingly identical doped materials, as well as between different labs, because of the sensitivity of PLQE to non-radiative losses. While we do indeed observe strong PLQE increases and show that they are also concomitant with a reduction in non-radiative recombination rate (see Supporting Fig. S4), it is thus the doping-induced increase of the intrinsic *radiative* rate that is most remarkable and more fundamental to compare here. It is also unaffected by the varying (trap-density-convoluted) PLQE values reported for identical compositions by different labs. Aside from allowing operation at higher maximum brightness for LED applications in lighting and displays for a given semiconductor, an increased radiative rate will also enable important technologies based on quantum coherent phenomena, like those relying on lasing[16], single-photon emitters[3] or superfluorescence[17,18].

We rationalize these experimental observations by performing electronic structure calculations using density functional theory (DFT, see Supporting Information for details). We model the undoped and doped $CsPbX_3$ systems for different halide environments (@X = Cl, Br) using 3×3×3 supercells, corresponding to nominal doping concentrations of 3.7%, to accurately reflect the low doping concentrations in the synthesized materials.

In Figure 1c, we show that the optical bandgap increases upon doping compared to the pristine perovskite. In a previous study we showed this result for the specific case of Mn doping and could relate it to the concept of local perovskite lattice periodicity breaking through the dopant.[10] Here, we expand this observation to be valid also for all other doped systems studied and explain how any dopant will exhibit the same effect. Moreover, as shown experimentally above (Fig. 1b), the electron-hole overlap increases upon doping, and with it the radiative recombination rate – an intrinsic property that is very hard to influence generally for a given semiconductor. Within our theoretical framework this can be approximately tracked as the effective carrier mass of the electronic transition (Fig. 1d). We find that the effective mass increases significantly for all studied dopants, by 48-76% compared to the undoped system, and reproduces the measured radiative rate trends remarkably well. In detail, the observed increase for a chloride environment (@Cl) is strongest for Mn, followed by Zn and, only slightly smaller, Ni doping. The electron effective mass exhibits larger increases for the Mn@Cl case than for the other dopants, whereas for the bromide environment (@Br), Zn doping shows the largest effective mass increase.

We now rationalize all these observations, both experimentally and from a high level of theory, in a unified approach based on orbital symmetries, and demonstrate its implications for the charge carrier distribution in real space for the different cases, as shown in Figure 2.

We show that the dominant driving force for the bandgap increase as well as importantly the effective mass and radiative rate gains in all systems is the doping-induced lattice periodicity breaking, which is *element-unspecific*. Therefore, every doped material (Fig. 2b,c) exhibits a distortion in the charge density distribution compared to the homogeneously delocalized one observed for the pristine perovskite (Fig. 2a). Following the orbital model that has been proposed by Goesten and Hofmann[19] for pristine perovskites before, the cubic perovskite conduction band is constructed by an antibonding atomic $PbX_3$ basis (Fig. 2d, see Supporting Information and Figs. S11-13 for a more detailed discussion). To obtain delocalized bands, a phase factor is introduced at different wave vectors k in the electronic Brillouin Zone (BZ). At the R-point [k = (1/2,1/2,1/2)], the Bloch state exhibits a phase change between each neighbouring antibonding $PbX_3$ orbital basis, creating a bonding interaction between them (see also Supporting Figs. S6-7) and consequently stabilizing the electronic state at the R-point (thus becoming the conduction band minimum, CBM). This universally explains the blue-shift upon doping.

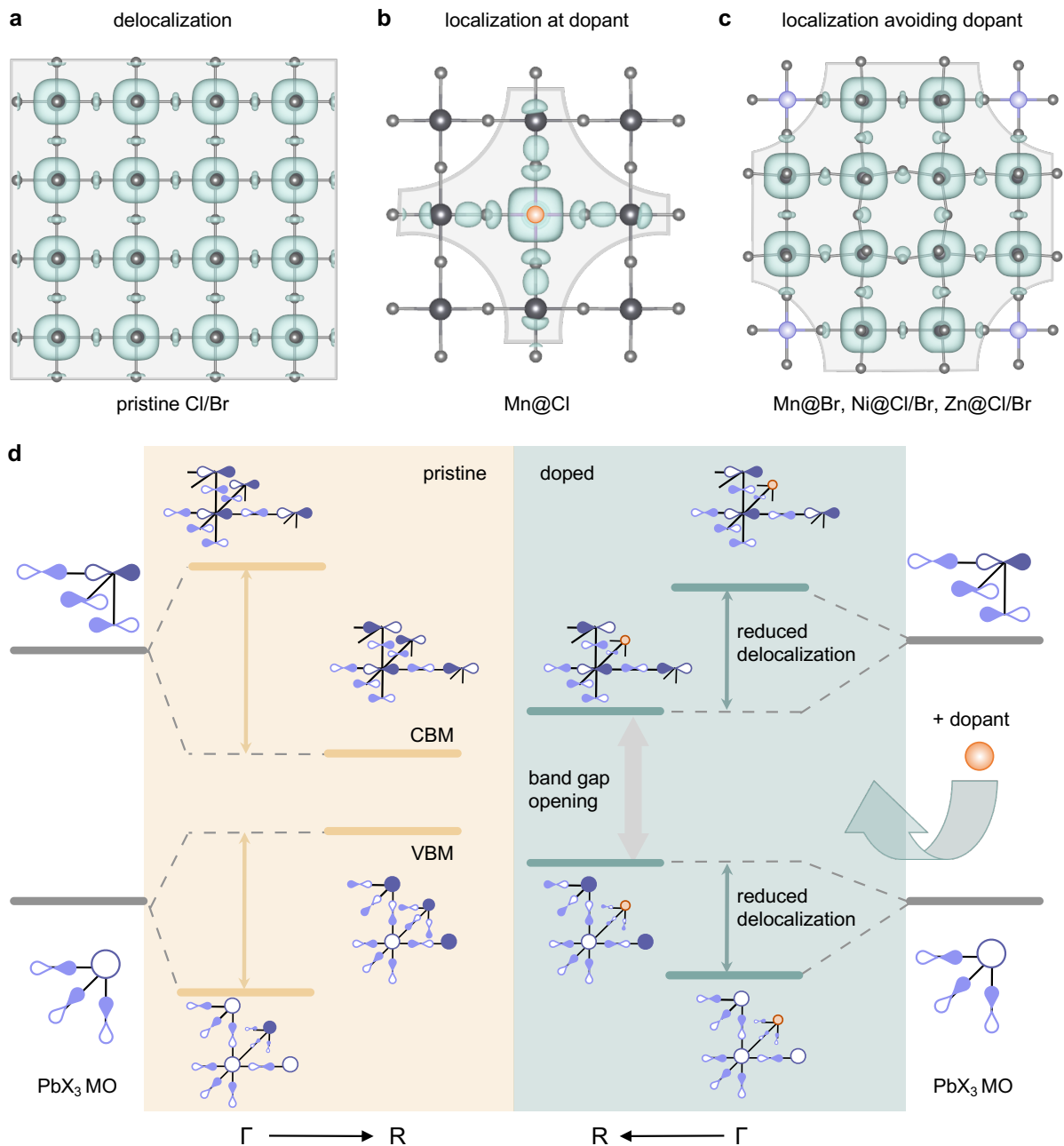

**Fig. 2 | Dopant- and halide-environment-dependent localization mechanism in perovskite nanocrystals. a**, Isosurface of the conduction band minimum electron density (green), delocalized across the lattice (grey) in undoped perovskite NCs. **b**, Charge localization induced by the dopant in the special case of Mn (orange) in a chloride environment, where hybridization occurs. Value at electron density isosurface is $2.36 \times 10^{-3}$ $e$ Å$^{-3}$. **c**, Reduced charge delocalization through dopant potential well formation in absence of hybridization, as is the case for all other dopant-halide systems studied (here shown for the example of Zn, blue, in a chloride environment). All cases result in faster radiative rates.

Results based on 3×3×3 supercells. **d**, Underlying orbital model. Left: Schematic representation of the Bloch states formed from the perovskite host PbX$_3$ molecular orbitals (MOs) at the Γ-point and the R-point, contributing to the valence and the conduction bands, respectively. The yellow double arrows indicate the bandwidth of the bands. Right: Representation of the same states upon incorporation of any B-site dopant (orange). It becomes clear that i) the band gap widens because of stabilization and destabilization of the VBM and the CBM, respectively (see grey double arrow), and ii) that the bandwidth for both valence and conduction bands is reduced upon doping (green double arrows), resulting in higher effective masses and radiative rates.

To explain the increase in effective masses (and radiative rates) observed for all dopants, we also need to consider the electronic CB states towards the BZ center (k = 0), *i.e.* not only at the R-point but also at the Γ-point. At Γ, the electronic band does not exhibit a phase change between the neighbouring PbX$_3$ bases, thus resulting in a destabilizing antibonding interaction between each site, in contrast to the bonding interaction at the R-point. Consequently, B-site doping leads to the opposite effect in the zone center compared to the zone boundary and stabilizes the conduction band towards Γ. An analogous argument leads to the converse effect for the valence band. Therefore, the simultaneous destabilization at the R-point and the stabilization at the Γ-point upon doping ultimately reduces the bandwidth and dispersion of both conduction and valence bands.

This reduced dispersion manifests itself in real-space as a *reduced delocalization* of the charge density with charge depletion at the B-site dopant which now acts as a potential well (Fig. 2c), resulting in *increased effective masses and radiative rates*, as we observe. As such, the electronic response is dictated by the symmetry of the atomic bases building the band edge states, and is therefore universal for all doped cubic lead-halide perovskites. To demonstrate this, we extended our analysis to the chemically dissimilar alkaline earth metals, and present an extensive study in the Supporting Information (Supporting Figs. S7-S10), confirming the effect of lattice periodicity breaking is largely independent of the electronic configuration of the perturbing dopant element, and mostly depending on its size (*i.e.* Goldschmidt factor, Supporting Fig. S8) and concentration (*i.e.* amount of perturbance, Supporting Fig. S9).

We also confirm the role of the *electronic effect* by which a dopant can affect the charge density distribution. This effect is *element-specific* and is found to be significant only in the case of manganese, and only in a chloride environment (Fig. 2b). Here, the s orbitals do match well energetically and symmetry-wise with the perovskite conduction band edge and thus show a significant degree of hybridization, absent in all other studied systems. These newly formed hybridized states are lower in energy than the pristine perovskite and therefore lead to a localization around the dopant, thus strongly enhancing the radiative rates. While Mn@Cl therefore provides a special case within the studied systems with respect to the mechanism, *i.e.* charge localization *toward* the dopant rather than away from it, as is the case for all other systems studied, the resulting reduced spread of the Bloch waves and concomitantly observed radiative rate increase is very similar.

However, manganese in a chloride-rich environment is also the only doped system where the dopant d orbitals lie energetically within the perovskite bandgap, as becomes evident in the project density of states for the different dopants (Fig. 3).

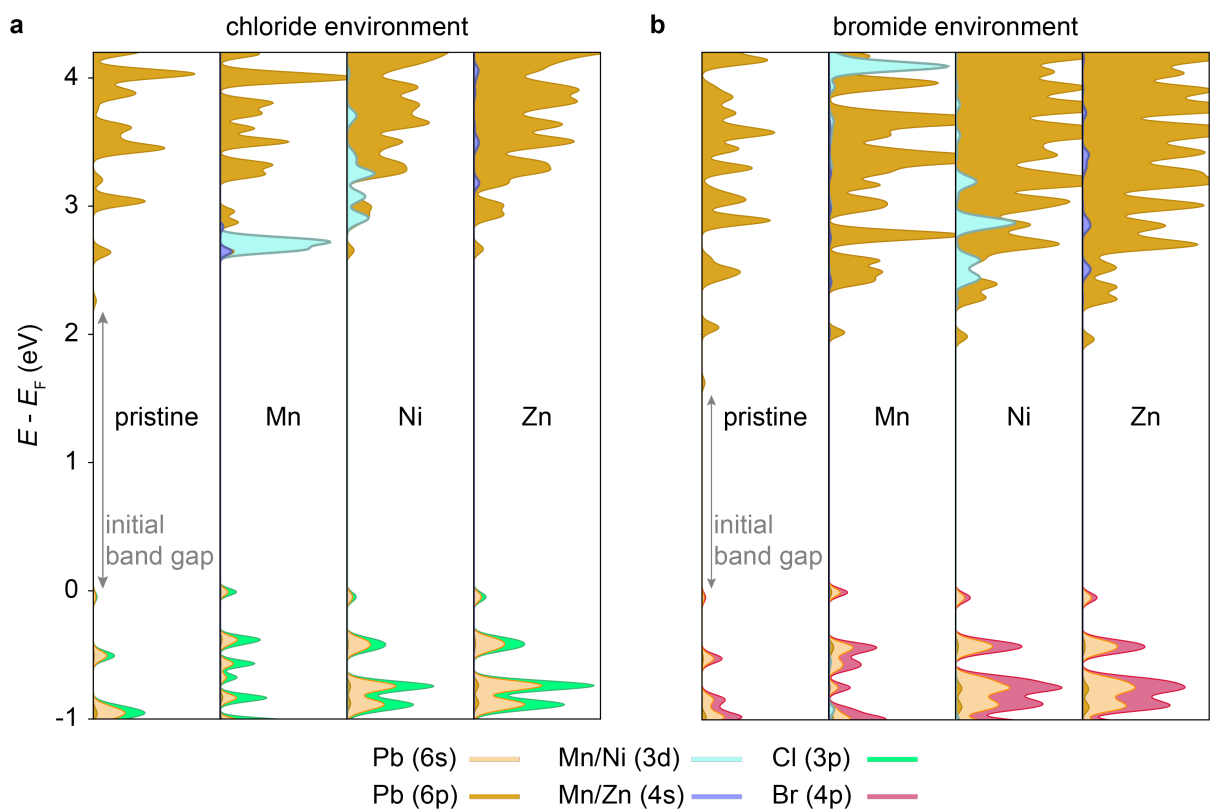

**Fig. 3 | Projected density of states (pDOS) of pristine and doped halide perovskites. a**, Dopant-dependent pDOS for a chloride environment where only the Mn 4s orbitals can hybridize with the perovskite conduction band edge states, leading to an enhanced localization effect (see Supporting Information for details). **b**, Dopant-dependent pDOS for a bromide environment where no dopant hybridization occurs. Grey arrows indicate initial band gap of pristine systems before doping-induced widening occurs. Results based on 3×3×3 supercells.

For Mn, these d states do not hybridize with the perovskite host band edges and instead act as the commonly observed energy loss channel for excitations, resulting in a spin- and symmetry-forbidden transition. This is commonly observed upon a threshold Mn doping concentration as long-lived orange (~600 nm) emission[20,21]. It limits the application of manganese doping to achieve color-pure and bright blue emission in LEDs: A maximal doping concentration of about 0.2% (Mn:Pb atomic ratio)[10] exploits the doping-induced radiative rate benefits without yet forming those additional loss channels in significant amounts which overall reduce the PLQE of the blue perovskite emission beyond this concentration.

In contrast to manganese in a chloride environment, for the other systems the dopant d orbitals lie energetically higher within the perovskite conduction bands and thus cannot act as such a loss channel. For Ni, there is, however, a significant DoS for the d orbitals present which partially screens charges and thus reduces the electron-hole attraction that leads to radiative recombination. Zinc with its closed d shell instead shows no significant density of d-orbital states and thus displays a slightly larger radiative rate compared to nickel (the same holds true for the closed-shell alkaline earth elements), as electron-electron correlations are less pronounced, and as is also confirmed experimentally (Fig. 1). For a bromide-rich environment the energy levels are shifting such that Mn s orbital hybridization becomes negligible and the additional electronic effect compared to the other dopants vanishes, as does the orange d-band emission. These findings have profound impact on the practical guidance of dopant choice for applications, as we will discuss below.

Lastly, we also performed ultrafast transient absorption (TA) spectroscopy to rule out other, more exotic doping-induced effects that one might hypothesize could play a significant role in the optoelectronic properties of these systems under device-relevant conditions and which would impact the choice of dopant; specifically multi-exciton and spin-related phenomena taking place at ultrafast time scales (Fig. 4).

For extracting bi-exciton lifetimes, we first compare the ultrafast transient absorption kinetics at a very low and a very high excitation density (Fig. 4a, <$N$>$_0$ is the initial average excitation density per NC, see Supporting Information for details). At high laser fluence, the formation of multi-exciton species can be observed which can lead to altered carrier dynamics and emission properties as discussed in the literature[22–24]. By subtracting the late-time normalized low- from the high-fluence data, the multiexciton recombination rate can be estimated from a monoexponential fit to the resulting early-time kinetics as shown before by Klimov and others[23] (Fig. 4b). We observe very similar multiexciton (*i.e.* most likely biexciton) recombination times of about $\tau_{BX} = 16 \pm 2$ ps for *all* studied doping systems and therefore rule out any significant doping-induced modification of multi-exciton dynamics.

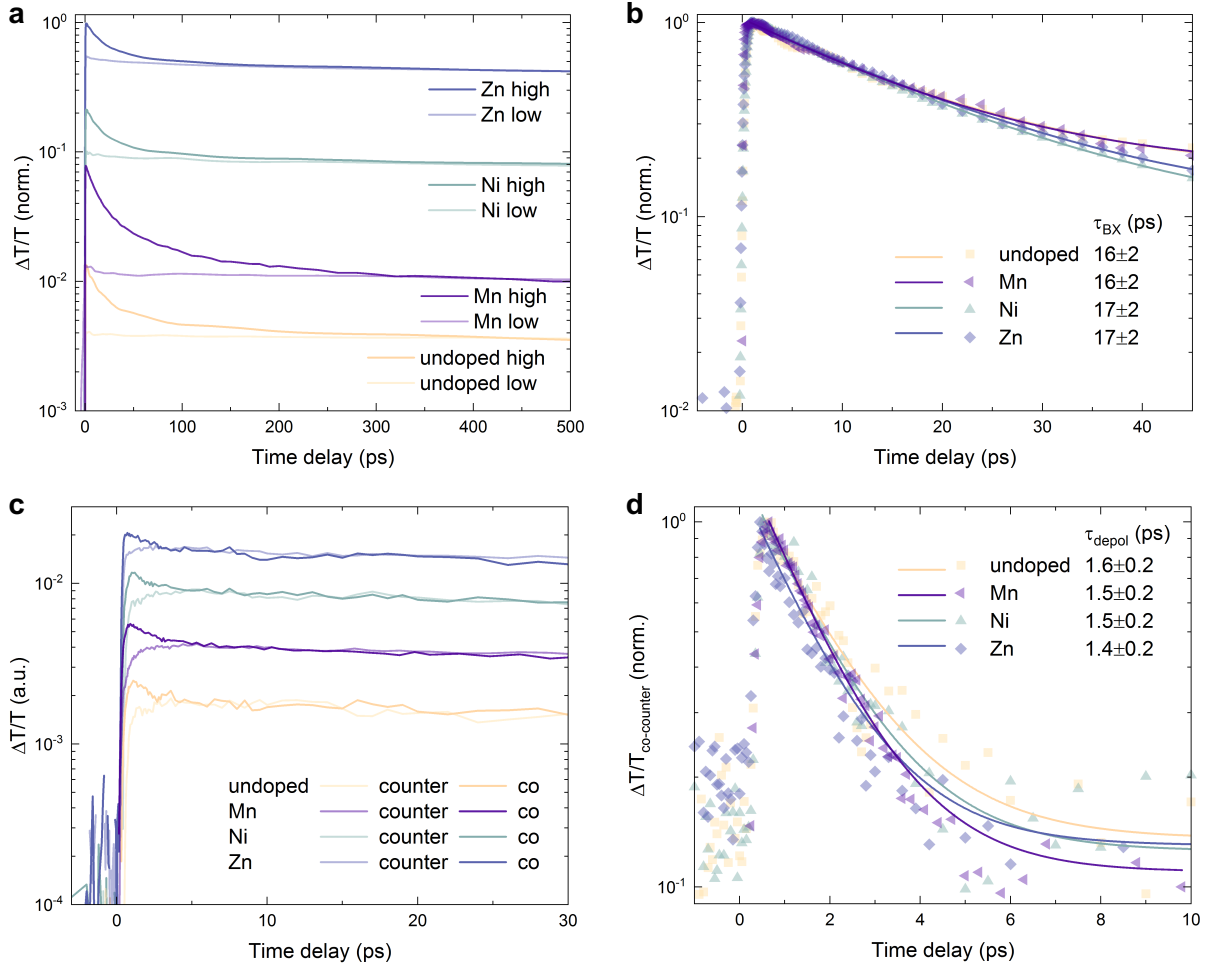

**Fig. 4 | Doping does not affect biexciton and spin lifetimes in NCs under device-relevant conditions. a**, High excitation-fluence (dark curves, $<N>_0 = 31$) and low-fluence (light curves, $<N>_0 = 0.1$) ground-state bleach kinetics extracted from transient absorption (TA) data, vertically off-set for clarity. **b**, Monoexponential fits to the kinetics obtained by subtracting the low- from the high-fluence data in **a**. Very similar biexciton lifetimes are extracted for all compositions. **c**, Co-polarized (dark curves) and counter-polarized (light curves) ground-state bleach kinetics extracted from TA data (fluence $<N>_0 = 0.3$, respectively), vertically off-set for clarity. **d**, Monoexponential fits to the kinetics obtained by subtracting the co- from the counter-polarized data in **c**. Very similar spin-depolarization lifetimes are extracted for all compositions. All samples were photoexcited at 400 nm (~100 fs pulses, repetition rate 1 kHz) and measurements performed at room temperature.

Finally, to determine the spin-relaxation times we studied the compositions using circularly polarized transient absorption spectroscopy (Fig. 4c,d). Since some of the transition-metal dopants possess a net magnetic moment due to their unpaired d-electrons, one might expect an influence on the spin-dynamics in the doped perovskite systems[25]. We compare the co- and counter-polarized kinetics and (Fig. 4c) and subtract them to retrieve the spin-depolarization lifetime through a monoexponential fit of the resulting decay[4,25,26] (Fig. 4d). We find that this lifetime, at which an equilibrium between the different total angular momentum sub-states ($m_J$ = -1/2 and +1/2) is reached, is very similar across *all* doped systems and estimated to be $\tau_{depol} = 1.5 \pm 0.2$ ps. We therefore also exclude any significant doping-induced influence on the spin dynamics of the perovskite with respect to the optoelectronic properties at room temperature, leaving the discussed charge localization effects as the only relevant ones that intrinsically impact the light emission under device-relevant conditions. While a recent report on the influence of nickel doping suggests a certain degree of spin-exchange coupling[27], those measurements relied on the presence of strong magnetic fields and low temperatures, making the phenomena conceptually interesting for spintronics applications but unlikely to impact the optoelectronic properties under device conditions, *i.e.* at room temperature and zero field. We can thus now conclude with the following remarks for the informed choice of dopant system.

**Guidelines arising for choice of dopant based on application**

It follows that, if the desired device emission is to be optimized for higher wavelengths, a bromide-rich halide environment is to be chosen over a chloride one due to the lower bandgap of the former. Here, we recommend the choice of zinc as dopant, as i) no additional s-orbital hybridization could be exploited in this environment anyways, ii) $Zn^{2+}$ then shows the strongest dopant-induced lattice periodicity breaking effect for enhancing the radiative rate, without the partial screening present for the open-shell configurations in $Mn^{2+}$ and $Ni^{2+}$, while iii) also avoiding potential d orbital induced loss channels present in $Mn^{2+}$. If the desired application instead calls for a broad, dual, or long-lived emission spectrum, manganese doping with its additional orange (600 nm) d-state emission should be chosen, which is also not present in the partially filled d orbitals for nickel, due to its energetic position. If the device is instead meant to be used in spintronics applications, then both Mn and Ni could be potentially useful due to their unpaired spins. If the desired emission wavelength should be in the very blue spectral range instead, a chloride-rich halide environment is to be used in general. As such, in the low-doping regime, manganese is the optimal dopant, as its s orbital hybridization leads to the strongest carrier localization, boosting the radiative rate gains further (and more so than the partial d-electron screening would counteract this, also since the DoS here is sufficiently far away energetically from the perovskite band edges). In the high doping regime, Mn becomes, however, suboptimal due to the formed d-state loss channels for the blue host emission. Thus, Zn would in this regime be the best choice (over Ni for the above electron screening reasons), or indeed alkaline earth metals. However, Ni & Zn form also less stable compositions, as measured by the PLQE dropping by about 30% within five days outside an inert atmosphere, while instead Mn-doped NCs sustained their PLQE for more than three months outside an inert atmosphere. This superior stability of Mn-doped NCs may be rationalized by the unique electronic effect of orbital hybridization in this case, as discussed above. Here the charges also become localized around the dopant for efficient radiative recombination (Fig. 2b), and they are thus less likely to diffuse to the trap sites distributed across the remaining perovskite lattice. In contrast, for the other dopant systems, the charge density becomes localized in the perovskite host by avoidance of the perturbing dopant (*i.e.* a reduction in delocalization, see Fig. 2c), and thus are more susceptible to the NC trap density. A balance between absolute performance gains and stability should thus be considered when choosing the correct dopant system for the desired application.

**Conclusion**

In summary, we have experimentally and theoretically established a generalizable model for the working mechanism of doping in perovskite nanocrystals. We found this to be a largely dopant-independent lattice periodicity breaking effect increasing radiative rates, which is further modulated by more subtle effects like orbital hybridization, screening from electron-electron interactions and respective halide environment. Our findings allow for the informed choice of the optimal doping system for a given optoelectronic device application.

Nanocrystals Using Optically Detected Magnetic Resonance Spectroscopy . *Chem. Mater.* (2022). doi:10.1021/acs.chemmater.1c03822

**Data availability**

The data that support the plots within this paper and other findings of this study are available at the following Online Repository: https://doi.org/XXXXX.

**Notes**

The authors declare no competing financial interest.

**Acknowledgements**

S.F. acknowledges funding from the Engineering and Physical Sciences Research Council (EPSRC UK) *via* an EPSRC Doctoral Prize Fellowship, and support from the Rowland Fellowship at the Rowland Institute at Harvard University. S.F., I.B. and B.M. are grateful for support from the Winton Programme for the Physics of Sustainability. Y.L. acknowledges the funding from Simons Foundation (Grant 601946). B.M. also acknowledges support from a UKRI Future Leaders Fellowship (Grant No. MR/V023926/1) and from the Gianna Angelopoulos Programme for Science, Technology, and Innovation. The calculations were performed using resources provided by the Cambridge Service for Data Driven Discovery (CSD3) operated by the University of Cambridge Research Computing Service (www.csd3.cam.ac.uk), provided by Dell EMC and Intel using Tier-2 funding from the EPSRC (capital grant EP/T022159/1), and DiRAC funding from the Science and Technology Facilities Council (www.dirac.ac.uk). G.H.A. and D.N.C. acknowledge funding from Stanford University. P.N. acknowledges the support of a Stanford Graduate Fellowship in Science & Engineering as a Gabilan Fellow. M.S.F. acknowledges the support of the Stanford Graduate Fellowship in Science and Engineering as a P. Michael Farmwald Fellow and of the National GEM Consortium as a GEM Fellow.

Supplementary Information for

# Universal mechanism of luminescence enhancement in doped perovskite nanocrystals from symmetry analysis

Ghada H. Ahmed[1], Yun Liu[2], Ivona Bravić[2], Xejay Ng[2], Ina Heckelmann[2], Pournima Narayanan[1], Martin S. Fernández[1], Bartomeu Monserrat[2,3], Daniel N. Congreve[1], Sascha Feldmann[2,4]*

[1]Department of Electrical Engineering, Stanford University, Stanford, CA 94305, USA
[2]Cavendish Laboratory, University of Cambridge, Cambridge, CB30HE, UK
[3]Department of Materials Science and Metallurgy, University of Cambridge, Cambridge, CB30FS, UK
[4]Rowland Institute, Harvard University, Cambridge, MA 02142, USA
*Email: sfeldmann@rowland.harvard.edu

## Methods

**Nanocrystal synthesis**

Precursor materials:
Cesium carbonate ($Cs_2CO_3$, 99.995%, metal basis), 1-octadecene (ODE, technical grade 90%), Lead (II) chloride ($PbCl_2$, powder 98%), Lead (II) bromide ($PbBr_2$, powder 99.99%), Oleylamine (OAm, technical grade 70%), Oleic acid (OA, technical grade 70%), Trioctylphosphine (TOP, technical grade 90%), Nickel(II) chloride (anhydrous, 99.99%), Zinc (II) chloride (anhydrous, 99%), Hexanes (98%, pure) all purchased from Sigma-Aldrich. All chemicals were used as received without further purification.

Preparation of cesium-oleate precursor:
Typically, $Cs_2CO_3$ (0.814 g) and ODE (40 mL) were added into a 100 mL 3-neck round-bottomed flask and dried under vacuum at 120°C for 1 hour. After degassing, the solution was heated under $N_2$ atmosphere. Then, (2.5 ml) of oleic acid (OA) was subsequently injected into the solution mixture, and heated under $N_2$ at 150°C until $Cs_2CO_3$ was completely dissolved yielding a clear solution. Afterwards, the solution was cooled to room temperature for storage, and reheated to 100°C before use, as the cs-oleate precipitates out at room temperature.

Synthesis of undoped $CsPb(Cl/Br)_3$ NCs:
165 mg of $PbBr_2$ (0.450 mmol), 83.6 mg of $PbCl_2$ (0.301 mmol), 20 mL of octadecene, 2 mL of dried oleylamine, 2 mL of dried oleic acid, and 2 mL of trioctylphosphine were loaded into a 100-mL three-neck round-bottomed flask, dried under vacuum at 130°C for 45 min. The flask was then filled with $N_2$ and heated to 150°C for 10 minutes under rigorous stirring. Following that, the temperature was raised to 165°C under $N_2$ and kept at this temperature for 5 min. Then, 1.7 mL of pre-heated cesium oleate solution was quickly injected into the solution. After 60 seconds, the reaction was quenched by immediate immersion of the reaction flask into an ice-water bath.

Synthesis of $Zn^{2+}$ and $Ni^{2+}$ doped $CsPb(Cl/Br)_3$ NCs:

For a typical synthesis of doped $CsPb(Cl/Br)_3$ NCs, 0.21 mg (0.572 mmol) of lead (II) bromide, two different doping concentrations (0.051, 0.071, 0.081 and 0.091 mg) of $NiCl_2$ or $ZnCl_2$, and 0.080 mg $MnCl_2$ (for comparison purposes), 20 mL of octadecene, 2 mL of dried oleylamine, 2 mL of dried oleic acid, and 2 mL of trioctylphosphine into a 100-mL three-neck round-bottomed flask. This was dried at 130 °C for 45 min and heated to 150°C under vacuum for 10 minutes. The yielded solution was then heated to 165°C under $N_2$ protection for 5 minutes, after which 1.7 mL of pre-heated Cs-oleate precursors was rapidly injected into the solution. After having reacted for 60 s, a crude product was cooled to room temperature in an ice-water bath.

Post-synthetic halide exchange:

A blue shifted PL emission was observed after doping the NCs with both $Zn^{2+}$ and $Ni^{2+}$ ions as compared to the undoped ones, and it was more prominent for the $Ni^{2+}$ doped samples. Therefore, for comparison purposes, the PL was aligned back to 470 nm for all the NCs. The as-synthesized NCs crude solution was aligned by exchanging the nanocrystal solution with (1-5 ml) of $PbBr_2$ stock solution prepared as below:

2.2 g of $PbBr_2$ (6 mmol), 50 mL of octadecene, 2 mL of dried oleylamine, and 2 mL of dried oleic acid in a 100-mL round-bottomed flask at 130°C were all mixed and stirred under vacuum for 40 min, after which it was cooled down to room temperature. The amount of $PbBr_2$ stock solution added to the crude solution was carefully monitored by tracking the PL peak position before purifying or washing the nanocrystals.

Isolation and purification of the NCs:

After the crude solution was cooled with an ice-water bath, the aggregated NCs were separated by centrifuging for 5 min at 12000 rpm. Then, the NCs were redispersed in 6 mL of hexane and centrifuged again for 5 min at 12000 rpm, and the supernatant was discarded. After repeating the previous step one more time, the final precipitate was redispersed in 6 mL of hexanes. It should be noted that different isolation and purification protocols have been tested and the PLQE was carefully monitored. For these samples, it was found that adding antisolvent (*i.e.* anhydrous ethyl acetate or methyl acetate) quenches the PLQE for both the Zn and Ni doped samples.

**Structural characterisation**

Transmission electron microscopy (TEM) measurements were performed using a FEI Tecnai G2 F20 X-TWIN Transmission Electron Microscope with a 200 kV operating voltage. TEM samples were prepared by dropping a dilute colloidal solution of NCs in hexane onto the carbon coated copper grids and dried under ambient conditions.

**Compositional characterisation**

The concentrations of lead, nickel and zinc ions were determined by using a ThermoFisher Scientific X-SERIES II Quadrupole inductively-coupled plasma mass spectrometer (ICP-MS). For ICP-MS analysis, nanocrystals were stirred overnight in nitric acid to ensure the complete dissolution of the metals into the acid. The measurements were repeated multiple times per composition to confirm a reliable doping concentration in each case.

**Steady-state absorption**

A Shimadzu UV-3600 Plus spectrophotometer was used to collect the steady-state absorbance spectra of samples, which uses a photomultiplier tube. The final data shown is corrected for by measuring the same cuvette with the solvent (hexane) only and subtracting this spectrum from the one with nanocrystals.

**Steady-state and time-resolved photoluminescence (PL)**

Steady-state and time-resolved PL spectra were recorded by a gated intensified CCD camera (Andor Star DH740 CCI-010) connected to a grating spectrometer (Andor SR303i). The pulsed output from a mode-locked Ti:sapphire optical amplifier (Spectra-Physics Solstice, 1.55 eV photon energy, 80 fs pulse width, 1 kHz repetition rate) was used to produce 400 nm excitation via second harmonic generation in a β-barium borate crystal. The iCCD gate (width 2 ns) was electronically stepped in 2 ns increments, relative to the pump pulse, to enable ns-temporal resolution of the PL decay. Faster (~100 ps resolved) kinetics were recorded using time-correlated single-photon counting (TCSPC) employing a Picoquant system at 405 nm excitation.

**Photoluminescence quantum efficiency (PLQE)**

PLQE data was collected using the method described by de Mello *et al.*[1]. Briefly, samples were positioned in an integrating sphere and excited at 400 nm, while the PL was collected with an Andor Shamrock spectrometer and Andor iDus CCD array. A corrected value is then determined by collecting the light from the sphere without a sample, without hitting the sample and with hitting the sample, respectively. Stated values were determined on triplicate samples which were each measured thrice, hence reporting the average of nine measurements for each composition.

**Transient absorption (TA) spectroscopy**

TA is a form of pump-probe spectroscopy which measures the spectrally resolved variation in absorption by a sample under photoexcitation by a pump source. By varying the pump-probe time delay, the carrier recombination kinetics of the sample can be investigated. The third harmonic of a pulsed Nd:YVO$_4$ laser (Picolo-AOT MoPa) was used as the pump beam (~1 ns pulse width, 500 Hz repetition rate, 355 nm) for the ns regime measurements. The probe spectrum was generated using a white light quasi-continuum generated through pumping a CaF$_2$ window with the 800 nm fundamental of a Ti:Sapphire amplifier (Spectra-Physics Solstice). A delay generator was used to electronically vary the pump-probe delay. For the short time fs-regime, the pump beam was the second harmonic (400 nm) generated by the 800 nm fundamental passing through a β-barium borate crystal. The transmitted probe and reference pulses were recorded with an NMOS linear image sensor (Hamamatsu S8381-1024Q) and processed by a customized PCI interface from Entwicklungsbüro Stresing. For the spin-depolarization measurements, superachromatic quarterwaveplates (Thorlabs) were added to both the pump and probe beam paths, as well as broadband linear polarizers (Thorlabs) before each of these, cleaning up the respective polarization prior to transformation into left- and right-handed circularly polarized light. For co-polarized the same and for counter-polarized the opposite handedness of pump and probe beams was recorded, respectively. In order to not obscure the spectral response of a slightly changing probe beam upon handed-ness change, the pump polarization was selected to be changed instead. A large pump beam spot size (~1000 μm effective beam diameter) compared to a small probe beam spot size (~300 µm) ensured a high signal stability and homogeneity. Polarization degrees >90% were confirmed for pump and probe prior to each measurement using an analyzing broadband linear polarizer (Thorlabs).

**Calculation of excitation recombination rates**

We first determine first the total excitation decay lifetime from time-resolved PL. It is found that the PL kinetics for all compositions can only be fitted to a high satisfying level if not less than the sum of three exponentials is used according to:

$$PL(t) = \sum_{i=1}^{n} a_i e^{-t/\tau_X^i} \quad \text{(S1)}$$

with $\tau_X^i$ as effective single-exciton lifetime (including both radiative and nonradiative contributions) and $a_i$ the relative fraction (in sum being 1) of NCs in the $i$-th out of $n$ sub-ensembles possessing this lifetime. The necessity of a triexponential fit is in line with other reports on similar perovskite nanocrystals, for example by the groups of Klimov[2] or Herz[3]. As described in the references, it is assumed that all nanocrystals measured in the ensemble possess the same radiative recombination constant $k_{r,X}$, which is the inverse of the intrinsic radiative lifetime $\langle \tau_{r,X} \rangle$, while the nonradiative recombination constant $k_{nr,X}$ can vary for sub-ensembles, *e.g.* due to different trap densities or identities. From weighting the individual sub-ensemble lifetimes via $a_i$, we can thus determine the average PL lifetime $\langle \tau_X \rangle$ and hence the average total recombination constant $\langle k_X \rangle = \langle \tau_X \rangle^{-1}$. The PLQE of each sub-ensemble is the ratio of its radiative to total recombination rate, and the total PLQE of the ensemble is therefore:

$$PLQE = \frac{\langle \tau_X \rangle}{\langle \tau_r \rangle} \quad \text{(S2)}$$

From measuring PLQE and total PL lifetimes, the intrinsic radiative lifetime, and hence radiative recombination constant, can be determined. Furthermore, the average non-radiative recombination rate can be calculated, since

$$k_{nr,X} = \langle k_X \rangle - k_{r,X} \quad \text{(S3)}$$

The measured composition-dependent values and extracted recombination constants are summarized in Supplementary Table T2.

**Calculation of average excitations per nanocrystal $\langle N \rangle$**

The excitation fluence per pulse $F_{ex}$ (in µJ cm$^{-2}$) is related to the laser power $P$ used for photoexcitation of the samples in solution, the repetition rate $R_{rep}$ and the beam spot radius $r_{ex}$ through:

$$F_{ex} = \frac{P}{R_{rep} \pi r_{ex}^2} \quad \text{(S4)}$$

The photon fluence per pulse $j$ (in cm$^{-2}$) is given by:

$$j = \frac{P \lambda \pi r_{ex}^2}{R_{rep} hc} \quad \text{(S5)}$$

where $\lambda$ is the excitation wavelength, $c$ is the speed of light in a vacuum, and $h$ is Planck's constant.

The absorbed photon density per pulse $j_{abs}$ (in cm$^{-3}$) in the cuvette of pathlength $l$ that is being absorbed can be inferred from measuring the UVvis absorbance $A$ at the excitation wavelength and reads:

$$j_{abs} = j \frac{A}{l} \tag{S6}$$

The density of NCs in solution $\rho_{NC}$ (in cm$^{-3}$) is related to the NC concentration $c_{NC}$ (in mg mL$^{-1}$) and the volume of a NC $V_{NC}$ and the weight density of the material $\rho_m$ via:

$$\rho_{NC} = \frac{c_{NC}}{V_{NC}\rho_{NC}} \tag{S7}$$

Therefore, the average excitations per NC $\langle N \rangle$ determined by this approach is:
$$\langle N \rangle = j_{abs}/\rho_{NC} \tag{S8}$$

**Calculation of wavefunction overlap of electrons and holes**
The wavefunction overlap $\Theta$ of electrons and holes is defined as:

$$\Theta = |\int \psi_e^*(r)\psi_h(r)dV|^2 \tag{S9}$$

Following an approach previously used for example by the groups of de Mello Donegá[4,5], or Kelley[6,7] and originally described by Efros and Rodina[8], the overlap integral is directly related to the oscillator strength and radiative rate via:

$$k_{r,X} = \frac{2e^4 n}{\pi\varepsilon_0 m_0^3 c^3}|F|^2 \frac{m_0^2 E_g E_p \Theta}{3e^2\hbar^2} \tag{S10}$$

where $e$ is the elementary charge, $n$ the refractive index, $\varepsilon_0$ is the vacuum permittivity, $c$ is the speed of light in a vacuum, $m_0$ is the free electron rest mass, $\hbar$ is the reduced Planck's constant, and $E_g$ is the energy of the optical transition.

$F = 3\varepsilon_m/(\varepsilon_s + 2\varepsilon_m)$ is the local field factor to account for the screening of the nanoparticle, assuming a random orientation of NCs with respect to the electric field of the interacting light. For simplicity, we modelled the NCs as spheres, with $\varepsilon_m$ and $\varepsilon_s$ being the dielectric constants of the medium and semiconductor, respectively. A more rigorous treatment of the electric field strength inside cubic nanocrystals can be found in ref.[9] Importantly, this will not influence the changes to the wavefunction overlap, since the same shape of NCs is maintained without and with doping, as checked with TEM. $E_p$ is the Kane energy, usually found for III-V semiconductors to be on the order of 20 eV[2], but notably very close to 40 eV for CsPbX$_3$ perovskites due to the different orbital contributions to the valence and conduction band edges here, independent of halide choice[9]. The formal derivation of Eq. S10 can be found in the early works by Efros[8,10], and later also discussed together with Bawendi[11].

Importantly, the measured increases in oscillator strength (here extracted from time-resolved PL & PLQE; for the inverse approach via transient absorption, see our detailed study on the manganese system where we show the equivalence of both measurements explicitly[12]) and thus also in the radiative rate *cannot* be explained by the only slight increase in bandgap and Kane energy upon doping. This is because the doping level in the percent-regime is too small to significantly alter the global properties of the semiconductor, while still an exciton will experience the presence of a dopant, given that the distance between the distributed ions is less than 3 nm and the exciton Bohr radius already approx. 2.5 nm (then followed by diffusion as well). Thus, the overlap integral increases significantly in order to account for the enhancement observed experimentally. This stronger overlap is a direct consequence of the lattice-periodicity breaking by the dopants which leads to the localization of charges, thereby increasing the radiative recombination rate of charges in their vicinity.

**First principles calculations**
Computational methods and parameters
Density functional theory (DFT) calculations were performed using the Vienna ab initio simulation package (VASP)[13,14]. The core-valence interaction was described using the projector-augmented wave (PAW) method[15,16], with 9 valence electrons for Cs ($5s^25p^66s^1$), 14 valence electrons for Pb ($5d^{10}6s^26p^2$), 7 valence electrons for Cl ($3s^23p^5$) and Br ($4s^24p^5$), 13 valence electrons for Mn ($3p^63d^54s^2$), 10 valence electrons for Ni ($3d^84s^2$), and 12 valence electrons for Zn ($3d^{10}4s^2$). Due to the presence of the heavy Pb atom, all electronic structure calculations except structural relaxation were done including spin-orbit coupling (SOC) effects, which are included perturbatively to the scalar-relativistic Hamiltonian[17].

We first performed geometry optimizations of the undoped cubic $CsPbBr_3$ and $CsPbCl_3$ where the atomic positions and lattice parameters were allowed to fully relax using the semi-local PBEsol functional[18]. The electronic wavefunctions were expanded in a planewave basis with an energy cut-off of 400 eV, the Brillouin-zone was sampled with a 6×6×6 Γ-centered Monkhorst-Pack[19] k-point grid and the Hellmann-Feynman force convergence threshold was 0.01 eV/Å.

For the doped structures we constructed a 3×3×3 supercell from the relaxed cubic unit cell of $CsPbBr_3$ and $CsPbCl_3$, and a single Pb atom was replaced with Ni/Zn atom, for a doping concentration of 3.7%. The supercells were then relaxed with the lattice parameters fixed, and only the atomic positions allowed to move. At this doping concentration, we found fully relaxing both lattice parameters and atomic positions results in negligible structural changes. For the supercell calculations we employed a commensurate k-point grid of 2×2×2. For Ni doping, a ferromagnetic spin configuration was assumed, and for Zn doping, a non-spin-polarized calculation was used. To obtain the band structure, we used the PBEsol functional with SOC effects. The effective masses were directly calculated from the band structure *via* a parabolic fit (R → Γ) using the formula $\frac{1}{m^*} = \frac{1}{\hbar^2}\frac{\partial^2 E}{\partial k^2}$. It should be noted that the curvature of the band strongly depends on the functional choice and that our results should only be taken as a qualitative measure, but do match the observed experimental relative trends very well.

For the projected density of states (PDOS) calculation, we used the PBE0 functional[20,21]. To reduce the computation load, a lower energy cut-off of 300 eV was used, which can still reproduce the electronic structure accurately[22]. The bandgap was extracted at the R point. Mn was independently studied in a previous work we have published in ref.[12] using the HSE06 functional that calculates a slightly lower bandgap for the pristine and the doped system, the exact computational parameters can be found in the SI therein. We employed a scissor shift to the PDOS such that the optical bandgap matches the results using the PBE0 functional for better comparison.

## Supplementary Data

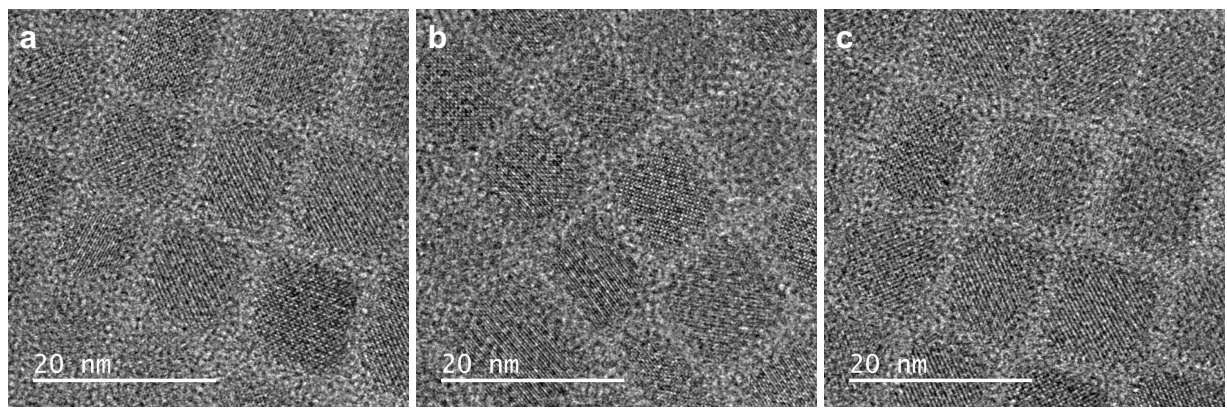

**Supporting Figure S1:** Scanning transmission electron microscopy images of pristine (a), nickel (b) and zinc (c) doped perovskite nanocrystals (NCs), respectively. See ref.[12] for manganese case. All undoped and doped NCs show a cubic morphology and an average crystal size of 10±2 nm, irrespective of composition, placing them in the weak quantum confinement regime. Scale bar is 20 nm, respectively.

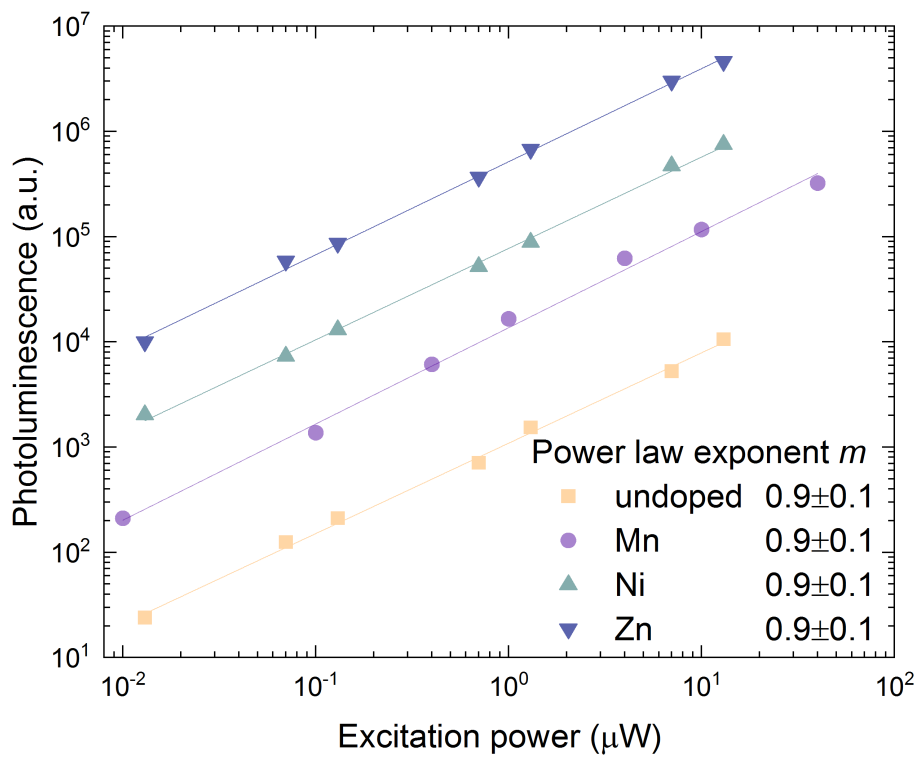

**Supporting Figure S2:** Excitation power dependence of photoluminescence intensity. Data traces are vertically off-set for clarity. The fit according to $P^m$ confirms a linear PL dependence on excitation power ($P$), expected for excitonic recombination.

| PbBr$_2$ (g) | PbCl$_2$ (g) | NiCl$_2$ (g) | PbBr$_2$ (mmol) | PbCl$_2$ (mmol) | NiCl$_2$ (mmol) | Molar ratio Ni/Pb | ICP-MS Ni$^{2+}$ (%) |
|---|---|---|---|---|---|---|---|
| 0.165 | 0.0836 | 0 | 0.450 | 0.301 | 0 | 0 | 0 |
| 0.21 | 0 | 0.071 | 0.572 | 0 | 0.547 | 0.956 | 1.10% |
| 0.21 | 0 | 0.081 | 0.572 | 0 | 0.625 | 1.092 | 2.01% |

| PbBr$_2$ (g) | PbCl$_2$ (g) | ZnCl$_2$ (g) | PbBr$_2$ (mmol) | PbCl$_2$ (mmol) | ZnCl$_2$ (mmol) | Molar ratio Zn/Pb | ICP-MS Zn$^{2+}$ (%) |
|---|---|---|---|---|---|---|---|
| 0.165 | 0.0836 | 0 | 0.450 | 0.301 | 0 | 0 | 0 |
| 0.21 | 0 | 0.071 | 0.572 | 0 | 0.520 | 0.909 | 1.50% |
| 0.21 | 0 | 0.081 | 0.572 | 0 | 0.594 | 1.038 | 1.55% |

**Supporting Table T1**: Typical amounts of precursors used for the Ni$^{2+}$ and Zn$^{2+}$ doped nanocrystals and atomic doping levels measured in the final NC samples. For a detailed doping concentration dependence study on Mn$^{2+}$ see ref.[12].

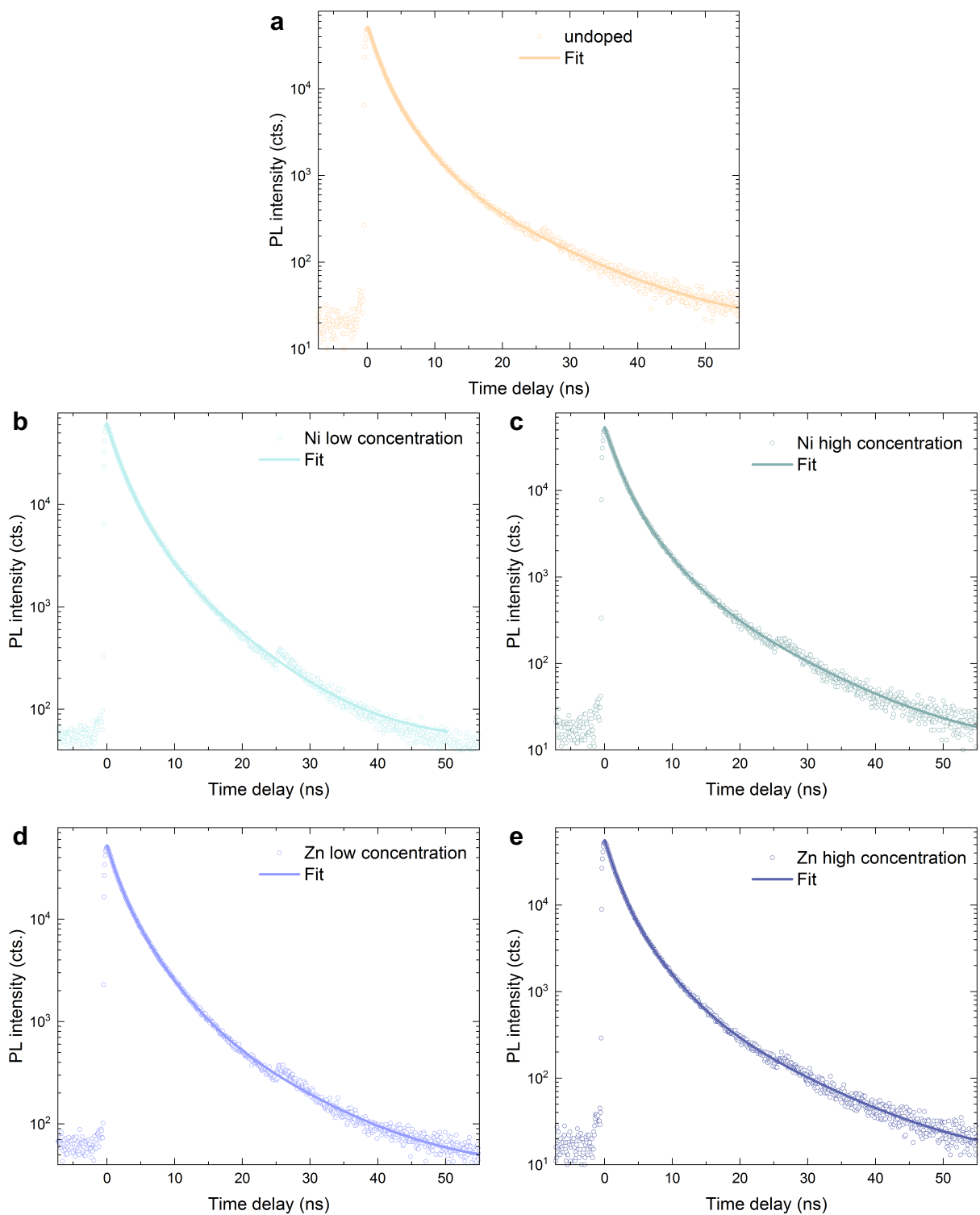

**Supporting Figure S3:** Time-resolved photoluminescence decays for **a**, undoped, **b**, 1% Ni doped, **c**, 2% Ni doped, **d**, 1% Zn doped, and **e**, 2% Zn doped nanocrystal dispersions in hexane. All samples excited with low-fluence 405 nm pulses (length ~500 ps) using the time-correlated single-photon counting method and integrated over the full spectrum. Bold lines show triexponential fits used for average lifetime determination for each composition (results summarized in table below). See ref.[12] for the full data set on Mn we reported earlier. The systematic small feature at 25 ns in all measurements is an instrument artefact due to pump

scatter, as confirmed by measurements on a dispersion of scattering titania particles showing the same response, and did not influence the fit parameters extracted.

| Composition | $\langle k_X \rangle$ ($10^8$ s$^{-1}$) [a] | PLQE (%) [a] | $k_{r,X}$ ($10^8$ s$^{-1}$) | $k_{nr,X}$ ($10^8$ s$^{-1}$) |
|---|---|---|---|---|
| undoped | 3.85 | 34 | 1.29 | 2.55 |
| Ni low | 3.71 | 41 | 1.51 | 2.20 |
| Ni high | 4.40 | 48 | 2.11 | 2.29 |
| Zn low | 3.66 | 44 | 1.63 | 2.02 |
| Zn high | 4.55 | 53 | 2.37 | 2.18 |

[a]Measurements taken at 405 nm excitation at a low-energy flux of 109 µW cm$^{-2}$.

**Supplementary Table T2.** Atomic doping simultaneously enhances radiative and decreases non-radiative recombination. The inverse of the average PL lifetime, as determined from TCPSC (Fig. S2), $\langle k_X \rangle$, and the non-radiative recombination constant, $k_{nr,X}$, are reduced upon doping, while the PLQE and the radiative recombination constant, $k_{r,X}$, increase upon doping, with zinc showing stronger improvements than nickel under otherwise identical conditions. $k_{r,X}$ was calculated according to $k_{r,X} = \text{PLQE} \times \langle k_X \rangle$ and $k_{nr,X}$ was calculated according to $k_{nr,X} = \langle k_X \rangle - k_{r,X}$, as discussed earlier, *e.g.* by Klimov and co-workers.[2] Similar results for manganese doping can be found in our earlier study focused on this dopant[12] and the linear increase in radiative rate with doping experimentally confirmed there was used for scaling the value shown in Fig. 1 to reflect the same doping level as for the other dopants.

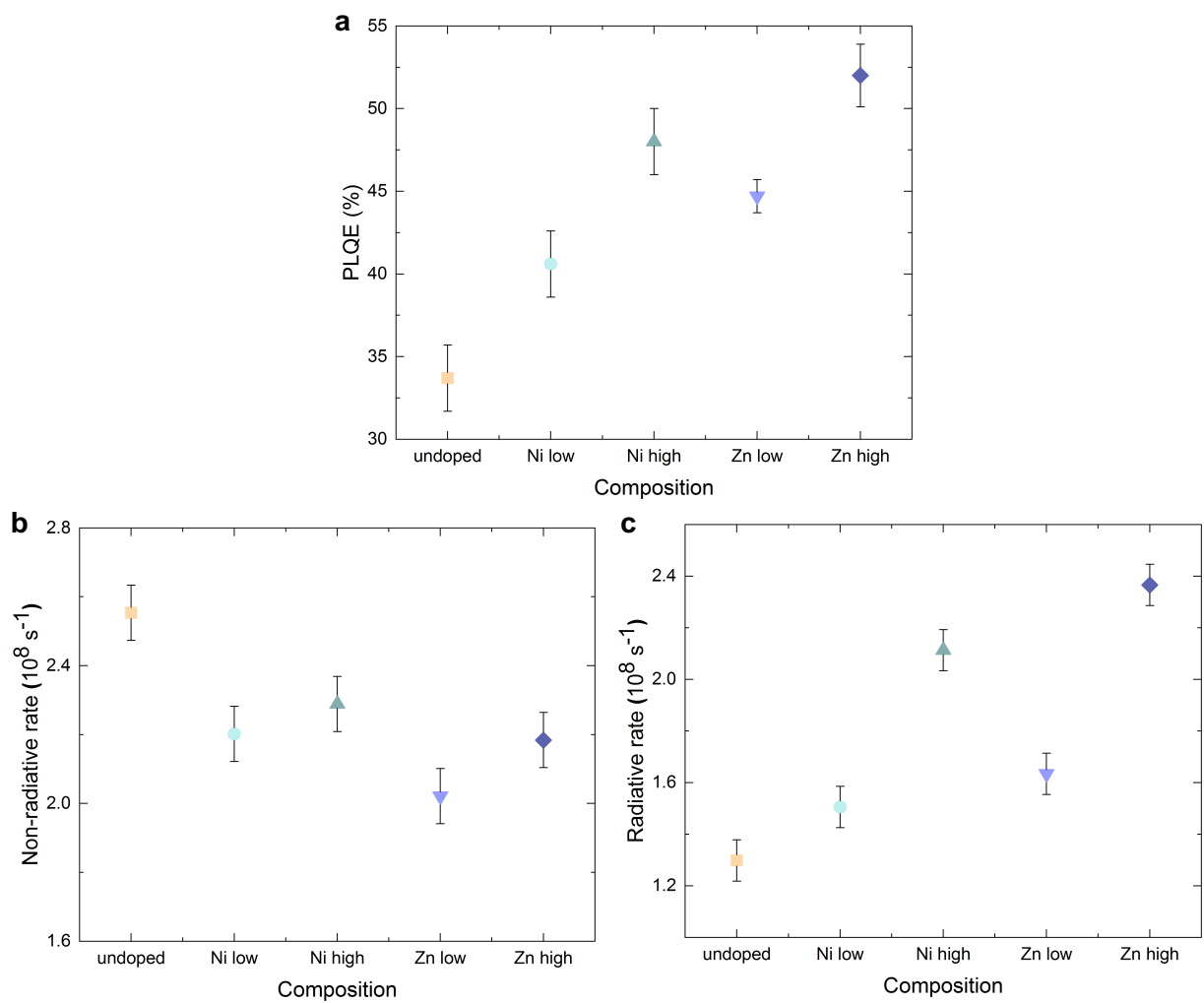

**Supporting Figure S4:** Concentration dependence of doping-induced changes to the NC emission properties. **a**, External PLQE, increasing upon doping with increasing doping concentration for Ni and Zn doping, respectively. **b**, Non-radiative recombination rate, which is reduced upon doping, and most efficiently so at low doping concentration, respectively. **c**, Radiative recombination rate, which increases upon doping, and increases with doping concentration. All samples were photoexcited at 405 nm at a low-energy flux of 109 μW cm$^{-2}$ (pulse length ~500 ps).

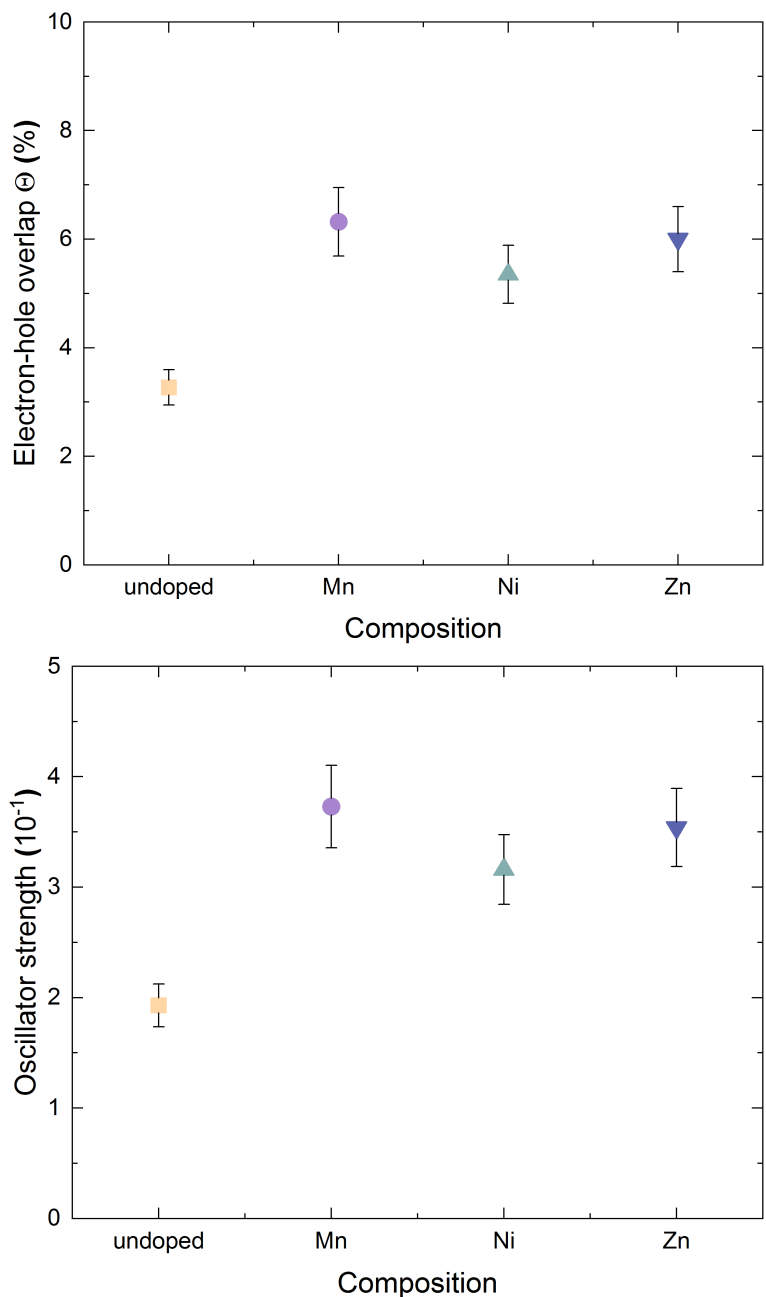

**Supporting Figure S5:** Overlap integral (top) and oscillator strength (bottom) extracted from experimental data (see Methods section for details). Both observables increase significantly upon doping due to lattice periodicity breaking.

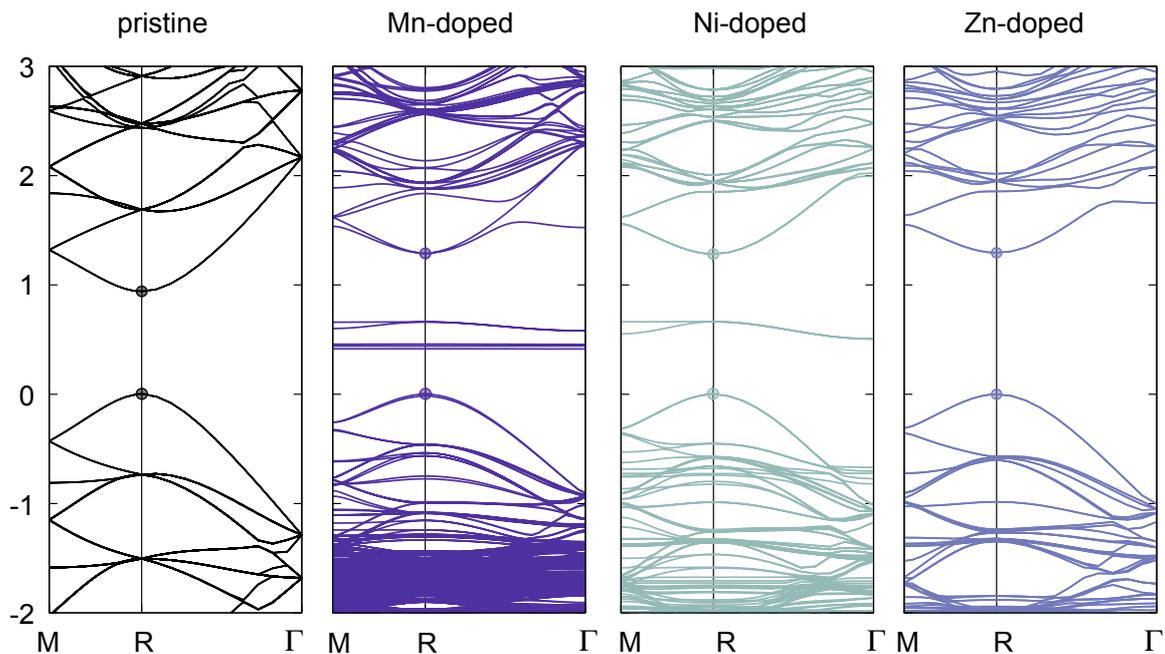

**Supporting Figure S6:** CsPbCl$_3$ band structures of the pristine and transition-metal doped perovskite (nominal doping concentration of 3.7%) calculated using the PBEsol functional with SOC. The effective masses were extracted using the perovskite band edges highlighted with circles. Black: band structure of pristine 3×3×3 CsPbCl$_3$ supercell with a direct band gap at the R-point. Purple: band structure of Mn-doped CsPbCl$_3$ with in-gap states corresponding to Mn d-states. We highlight that the position of the d-states is lifted towards the perovskite band edges when more accurate hybrid functionals are used in the main manuscript. Green: band structure of Ni-doped CsPbCl$_3$, again showing in-gap bands corresponding to dopant d-states. These are pushed into the conduction bands when PBE0 is used. Blue: band structure of Zn-doped CsPbCl$_3$.

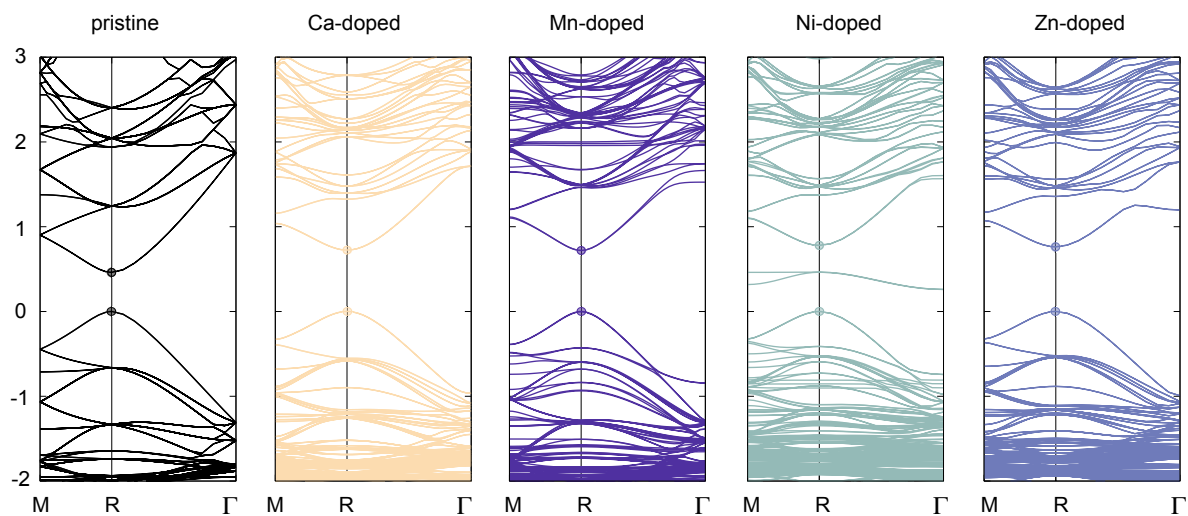

**Supporting Figure S7:** CsPbBr$_3$ band structures of the pristine and transition-metal doped perovskite (nominal doping concentration of 3.7%) calculated using the PBEsol functional with SOC. The circles highlight the band edges that were used to calculate the effective masses.

Black: band structure of pristine 3×3×3 CsPbBr$_3$ supercell with a direct band gap at the R-point. Yellow: band structure of Ca-doped CsPbBr$_3$. The Ca-doped effective masses were extracted as 0.144 (hole) and 0.15 (electron) electron masses, respectively. Purple: band structure of Mn-doped CsPbBr$_3$. Green: band structure of Ni-doped CsPbBr$_3$, again showing in-gap bands corresponding to dopant d-states. These are pushed into the conduction bands when PBE0 is used. Blue: band structure of Zn-doped CsPbBr$_3$.

<u>Study on alkaline-earth metal doping on CsPbBr$_3$</u>
To further study the role of ionic radii, concentration, and element-specific changes we conducted a thorough complementary first-principles study on the B-site doping of CsPbBr$_3$ with the alkaline earth (AE) metals beryllium (Be), magnesium (Mg), calcium (Ca), strontium (Sr) and barium (Ba). We have chosen the AE elements because they capture a wide range of ionic radii, electronegativities, electron affinities and all have unambiguous oxidation states of +2 that conserves the total charge of the crystal. Furthermore, due to the electron configuration of ns$^0$ we assume that both the filled and the empty dopant orbitals are well isolated from the perovskite band edges.

Again, all calculations are performed using density functional theory (DFT) with the projector augmented wave method as it is implemented in the Vienna Ab Initio Simulation Package. For the geometry optimisations we employed an energy cut-off 350 eV and a Γ-centred Brillouin zone (BZ) grid including 6×6×6 *k*-points for the primitive cell and commensurate grids for the supercells to relax the volume until the stress is below $10^{-2}$ GPa, while the internal atomic coordinates are fixed by symmetry. We perform all geometry optimisations using the generalised-gradient approximation of Perdew, Burke, and Ernzerhof (PBE) without the inclusion of spin-orbit coupling. We subsequently calculate the electronic properties using the same parameters but now we include the effect of spin-orbit coupling using the second variational method.

We start from a pristine primitive cell describing cubic CsPbBr$_3$. As an initial structural guess, we use the primitive cell as above and construct 2×2×2, 3×3×3 and 4×4×4 supercells. These supercell sizes give a sufficiently good range of doping concentrations for deducing trends but without reaching computationally prohibitive limits. We then introduce dopants into the pristine supercells through B-site substitution of one Pb atom. With the aforementioned range of supercell sizes, we are able to reach doping concentrations of up to 1.56%. In the first step we calculate the band gap of the doped systems without allowing the geometry to relax after substitution. This set of calculations allows us to deduce the effect of particle substitution of each AE atom on the band gap without any structural distortions arising from an ionic radius mismatch.

To understand how structural adaptation to the dopant changes the band gap, we then allow the internal coordinates to relax without changing the volume of the simulation cell. The atoms within the cell are allowed to move and twist to change the bond lengths and angles to accommodate for the new dopant.

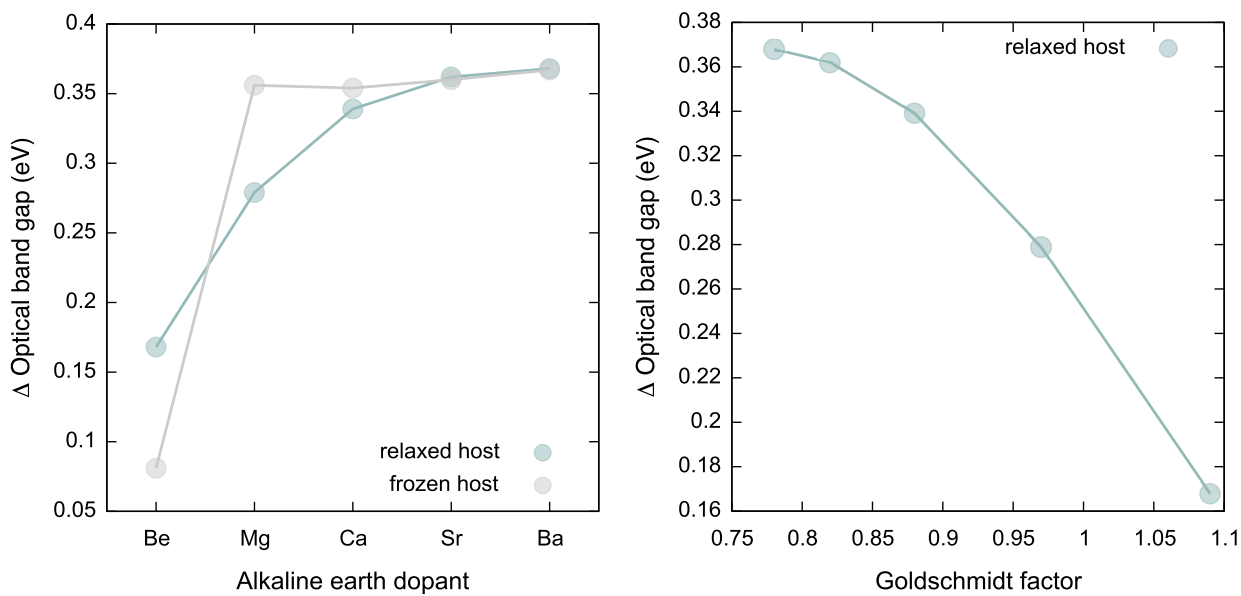

**Supporting Figure S8:** Dopant-dependent band-gap changes at a nominal doping concentration of 12.5%. Left, Calculated bandgap change in $CsPbBr_3$ upon doping with different alkaline earth (AE) metals. Grey: In this scenario, the replacement of a Pb atom with an AE atom is not followed by a structural relaxation. Green: In this scenario, the replacement of Pb with an AE atom is followed by a structural relaxation. At high concentrations, the structural adaptation of the perovskite scaffold leads to a blueshift with respect to the frozen configuration (with Be being the exception because it exhibits an in-gap state). Right, Change of the optical band gap against the Goldschmidt tolerance factor. The almost identical slope of the curve indicates that the size mismatch is the dominating factor that governs the difference between the frozen and the relaxed scenario.

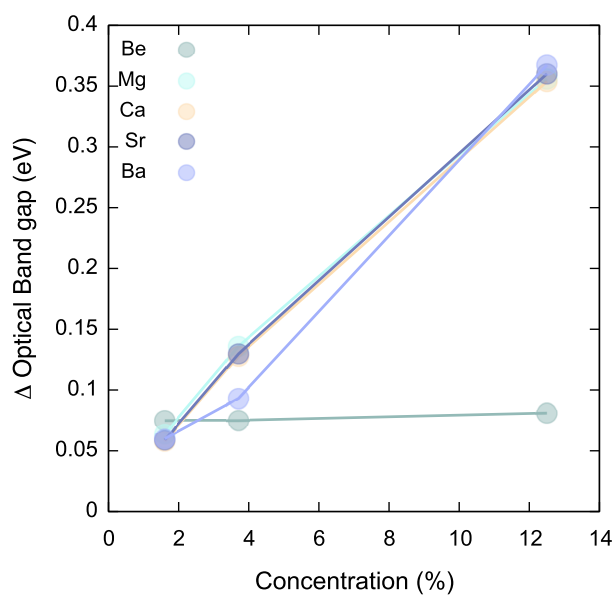

**Supporting Figure S9:** Calculated change of the band gap of $CsPbBr_3$ as a function of nominal doping concentration of all the alkaline earth dopants. In this case, the host structure is kept frozen. The concentration-dependent band-gap opening is linearly dependent on the nominal doping concentration (with Be being the exception because it exhibits an in-gap state), which is a result of breaking the periodicity of the Bloch states and pushing them towards the atomic limit.

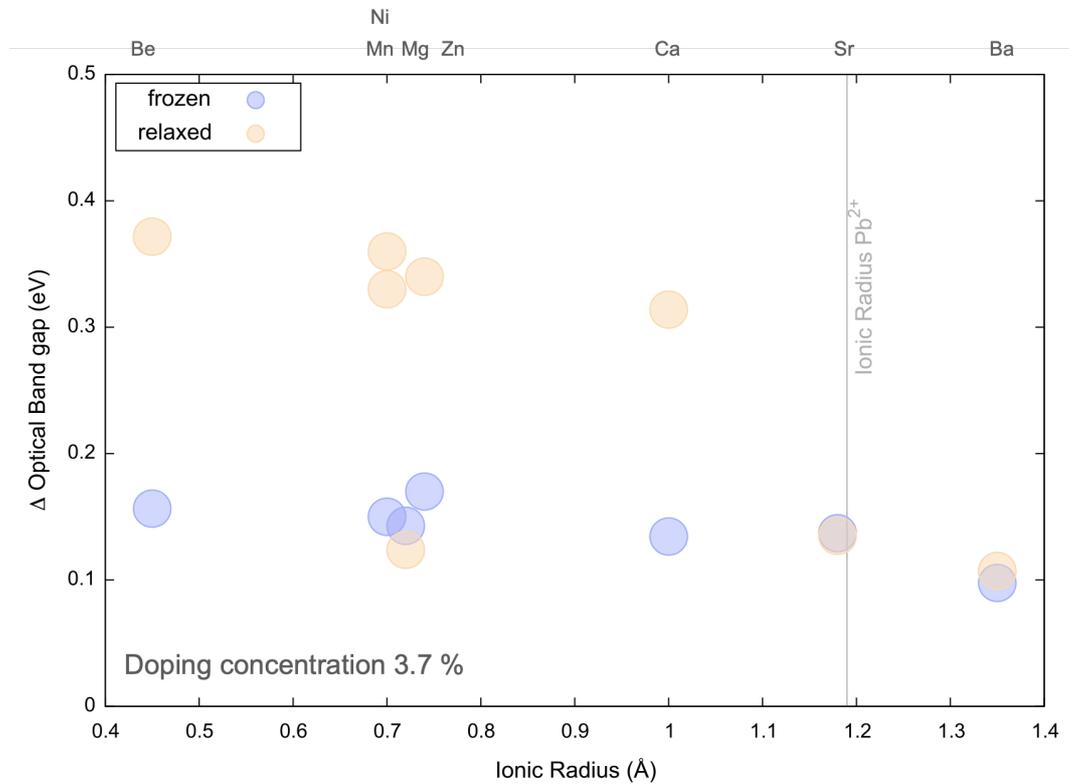

**Supporting Figure S10:** Change of the band gap as a function of the dopant ionic radius (Å) at low concentrations. Here, the optical band gap was calculated for the frozen as well as for the relaxed perovskite scaffold. It becomes evident that the structural relaxation of the host as a response to the B-site doping with a size-mismatched dopant (we added the transition metal results for completeness here) leads to a further band gap opening. This can be understood in terms of structural distortions that are captured in a bigger supercell but cannot be captured in small supercells. Here we see octahedral tilts in the perovskite scaffolds that are not present at higher concentrations (*i.e.* smaller supercells). Octahedral tilts in lead halide perovskites are known to widen the optical band gap due to the perturbed orbital interaction between the Pb and the halide atoms. This effect diminishes at strontium which has the same ionic radius as Pb and thus does not cause any structural changes due to ionic size mismatch.

Mechanistic origin of the band gap opening and effective mass increase

As highlighted in the main text, the response of the optoelectronic properties upon B-site doping (at low concentrations) follows very clear trends which we explain in a chemically intuitive fashion, bridging the gap between molecular orbital (MO) and electronic structure theory, using the linear combination of atomic orbitals (LCAO) ansatz and Bloch's theorem. For our discussion we closely follow the arguments of Goesten and Hoffmann. For the sake of simplicity, we do not include spin-orbit effects in this model as the symmetry arguments proposed here can also be used when spin-orbit effects are included.

We start with the smallest building block of the cubic perovskite structure as it is illustrated in Figure S11a. Here, the $PbX_3$ unit (Cs is omitted as it does not contribute to the relevant electronic states) is used to construct a cubic periodic lattice along the lattice vectors $a_1=a_2=a_3$, which are parallel to the Pb-X bonds of the $PbX_3$ unit. This leads to the cubic crystal structure. We also use the atomic $PbX_3$ unit to construct the near-band electronic states in the infinite lattice. For this, we show in Figure S11b the relevant atomic orbitals (AOs) that contribute to the electronic states close to the Fermi level, namely the Pb 4s and 4p, as well as the halide p.

For the $PbX_3$ unit with $C_{3v}$ symmetry we construct symmetry-adapted linear combinations of the halide p orbitals that create a sigma-type overlap with the Pb 4s orbital, creating a stabilized $\sigma$ fragment molecular orbital (FMO) with three bonding interactions, as well as a destabilized $\sigma^*$ FMO with three antibonding interactions. We analogously construct another set of FMOs, now using the Pb 4p and halide p AOs, which we refer to as the $\sigma\pi\pi$ basis. This symmetry-adapted linear combination leads to a threefold degenerate $\sigma\pi\pi$ and $\sigma^*\pi^*\pi^*$ basis. These FMOs do not contain the halide s orbitals but are sufficient for a qualitative description of the valence band (VB) and the conduction band (CB).

Using the FMOs we can build the delocalized Bloch states in the cubic perovskite crystal at the high-symmetry points Γ and R, according to Figure S12a. Key features of cubic lead halide perovskite which makes the electronic structure so chemically tangible are 1) that the lattice vectors are parallel to each Pb-X bond and 2) that the real and the reciprocal lattice vectors are parallel to each other (and hence the 1st BZ is cubic as well). Therefore, we can create the Bloch states for the zone center Γ and zone edge R along the real-space lattice vectors, now only including additional phases for different wavevectors k. We show the construction of the Bloch states along a reciprocal lattice direction below. For Γ, no phase change between different FMO units occurs, whereas at the R-point a phase change between each neighbouring FMO occurs. This picture, however, is yet incomplete and does not show us whether the generation of the Bloch states stabilizes or destabilizes the energy at the different high-symmetry points. For this it is necessary to consider the type of orbital interaction that occurs upon imposing translational symmetry. What happens with the different FMOs at Γ and R is depicted in Fig. S12b. For clarity, we only show the truncated $PbX_6$ octahedron. The $\sigma$ and $\sigma^*$ basis at Γ interact such that the octahedron exhibits three bonding and three antibonding interactions, making the states overall non-bonding and stabilizing them (the Bloch state corresponding to the $\sigma^*$ FMO is stabilized w.r.t. the FMO and the state corresponding to the $\sigma^*$ FMO is stabilized w.r.t. the FMO). In contrast, at R we see for the $\sigma$ and the $\sigma^*$ FMO six bonding and no antibonding, and six antibonding and no bonding interactions, respectively. This then suggests that the $\sigma^*$ states (which build the valence band) are most stabilized at the Γ point and most destabilized at the R-point (which is where the VBM is located). For the $\sigma\pi\pi$ and $\sigma^*\pi^*\pi^*$ basis the opposite is true and here we will only consider the $\sigma^*\pi^*\pi^*$ basis, since it builds the CB. At Γ, the state exhibits six antibonding interactions, destabilizing it w.r.t. the FMO, whereas at the R-point it only has three bonding and three antibonding interactions. These interactions result in a stabilization and destabilization of the Bloch state with respect to the FMOs at the Γ-point and the R-point, respectively.

Now that we understand how the translational symmetry stabilizes and destabilizes the electronic states along the Γ→R high-symmetry line, we can easily derive the effect of breaking translational symmetry by incorporating a B-site dopant.

In Fig. S13a we have illustrated the effect of doping on the band edges at R. Since the VB is the most antibonding at R (thus creation of a symmetry adapted linear combination through translational symmetry increases the energy of the Bloch state), breaking the periodicity will reduce the destabilization, pushing the VBM down in energy. The opposite is true for the CBM, where breaking the periodicity will push the CBM up because it perturbs the bonding interaction between the neighbouring FMOs. Both effects combined lead to an increase of the electronic band gap for every B-site dopant, as we have determined both experimentally and from first principles calculations.

This symmetry-encoded response to the dopant also explains the reduction of the effective charge carrier masses. Here, we consider the slope of the electronic bands along the R→Γ high-symmetry line and calculate the effective masses using a parabolic fit of the VB and the CB (which depend on the slope of the electronic band). Starting at the R-point of the VB, the state with the strongest antibonding interactions of the band, we move towards the zone center. Along that path, the strength of the antibonding interaction due to translational symmetry is continuously weakened until it reaches Γ, where the band is the most bonding. Breaking periodicity by incorporating a B-site dopant will cause the strongest stabilization effect at the R-point (since the translational symmetry induced antibonding interaction is the strongest at R), with the effect ceasing along the R→Γ high-symmetry line, up to a point, at which the bonding interaction dominates in proximity to the Γ-point (and thus breaking periodicity destabilizes the electronic states close to the zone center). This effect is illustrated in Fig. S13b. Considering the conduction band, the exact opposite effect occurs: Breaking periodicity causes the CB to destabilize most at the R-point and to progressively destabilize less, when we move along the R→Γ high-symmetry line. The effect on both the valence and conduction band leads to a change in the slope, which we qualitatively quantify using the bandwidth, and consequently to a reduced effective mass of charge carriers (*i.e.* reduced delocalization). We note that full first principles calculations show that the R→Γ high-symmetry line may cross with defect levels, but this should not affect the conclusions drawn here about the bandwidth.

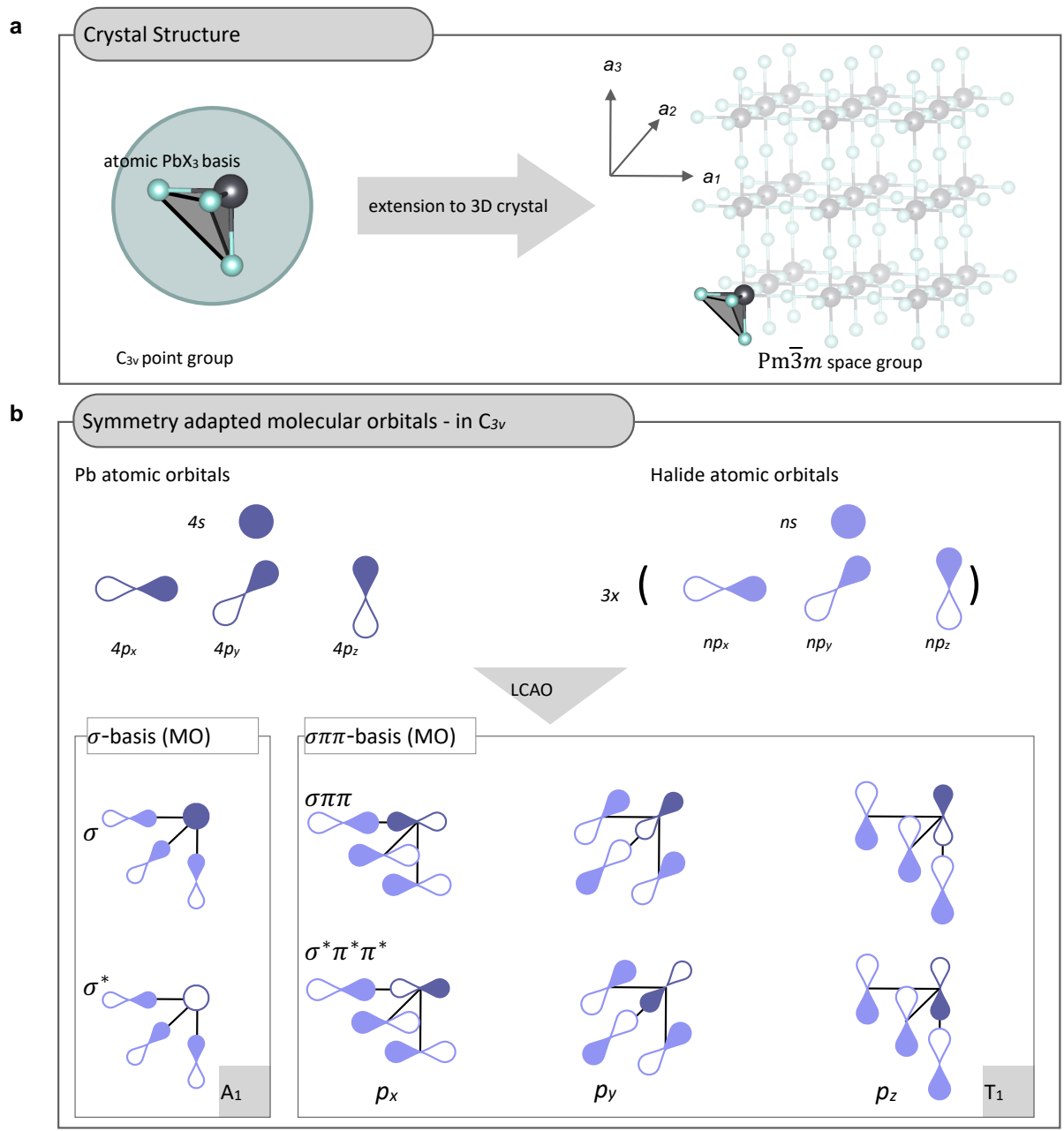

**Supporting Figure S11: a,** Illustration of the PbX$_3$ atomic basis with the point group C$_{3v}$ that is used to build the cubic perovskite crystal lattice by translation of the basis along the cubic lattice vectors $|a_1|=|a_2|=|a_3|$. The Cs atom is omitted here since it only negligibly contributes to the formation of the bands close to the band gap. In this crystal, each Pb lies at the Wickoff position *1a* (0,0,0) and has a point group symmetry of O$_h$. **b,** Top, valence atomic orbitals of lead and halide atoms. Bottom left, linear combination of atomic orbitals (LCAO) generates bonding and antibonding molecular orbitals of the atomic PbX$_3$ unit with $a_1$ symmetry (within the C$_{3v}$ point group). Bottom right, bonding and antibonding molecular orbitals of the atomic PbX$_3$ unit with $t_1$ symmetry (within the C$_{3v}$ point group).

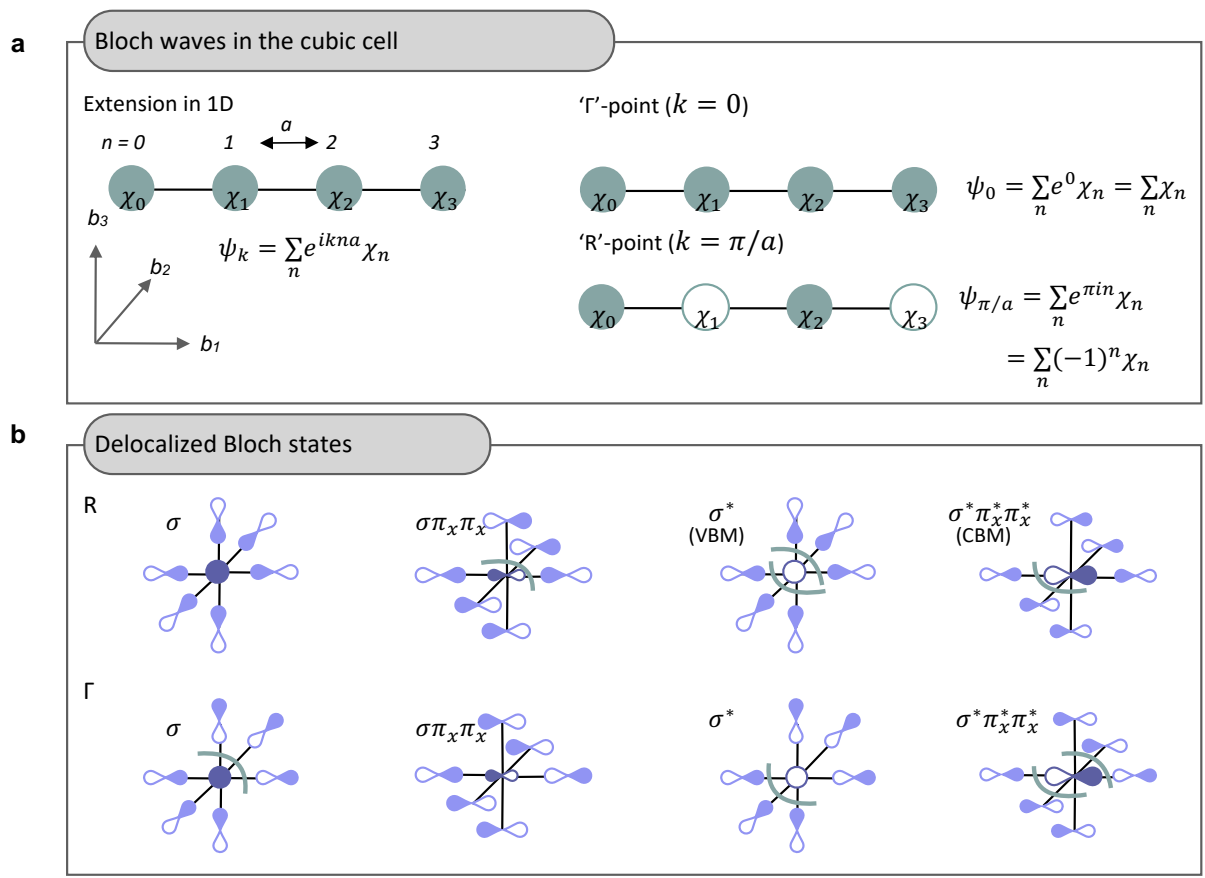

**Supporting Figure S12: a,** Formation of Bloch states along one of the identical reciprocal lattice vectors $b$ from any of the molecular bases of $PbX_3$ in a cubic lattice. At the Γ-point, there is no phase change between the repeating $PbX_3$ units, whereas at R the phase changes between each neighbouring $PbX_3$ unit. **b,** Formed Bloch states at the high-symmetry points Γ and R, truncated in order to illustrate the bonding and antibonding interaction as well as the symmetry of the isolated $PbX_6$ octahedral MOs. The green lines show the antibonding interactions between the Pb and the halide orbitals. The size of the orbital lobes is a qualitative measure of the energy-dependent orbital coefficients.

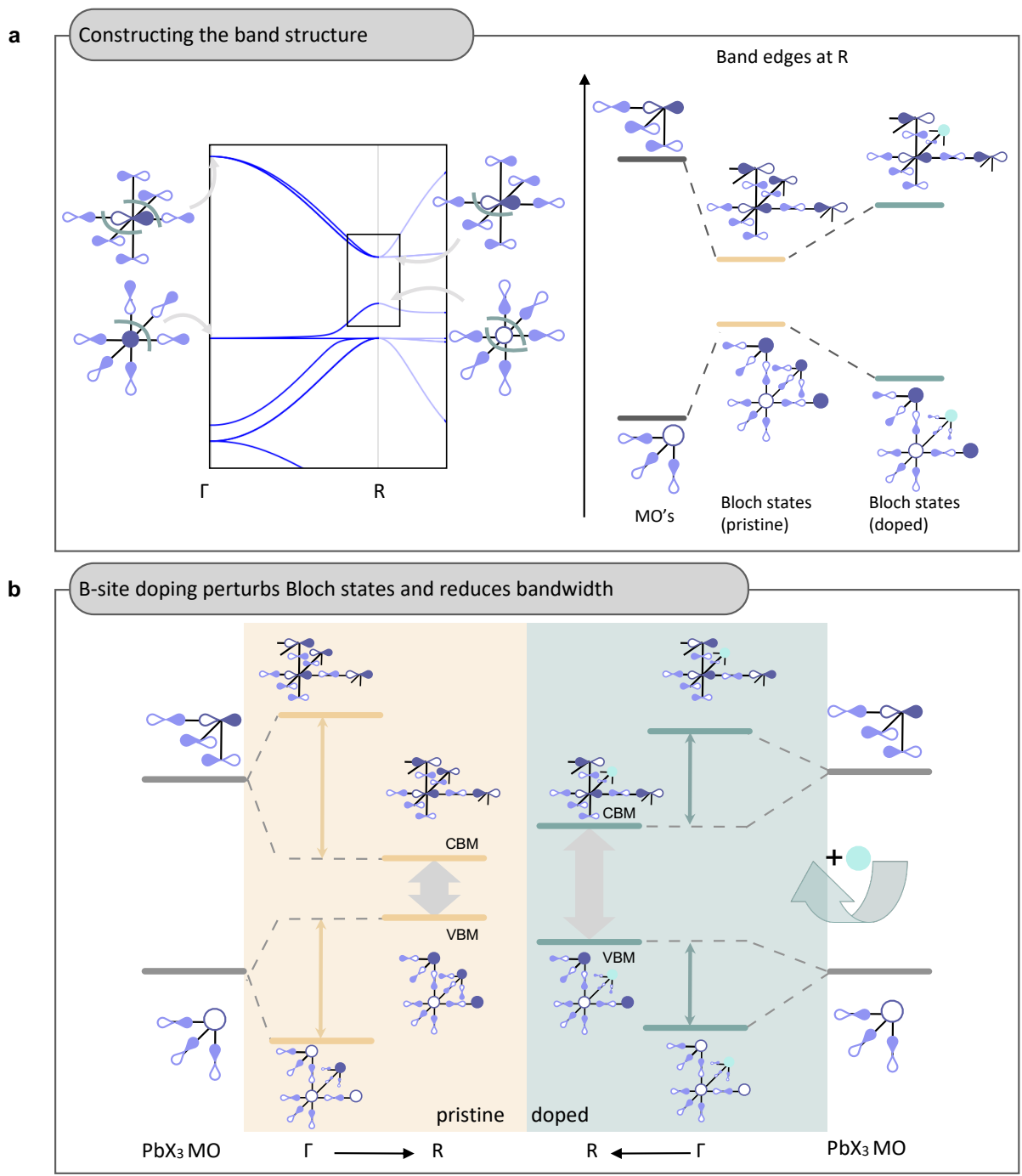

**Supporting Figure S13: a,** Left: Qualitative illustration of the band structure and the corresponding octahedral crystal orbitals for the VBM and CBM (both at the R-point), as well as the crystal orbitals of the lowest energy level in the valence band and the highest energy level in the conduction band (both at the Γ-point). Here it becomes clear that the bandwidth of the valence and conduction bands is determined by the bonding and antibonding nature of the Bloch states with different phase factors. Right: Schematic illustration of the band formation at the R-point starting from the molecular orbitals of the $PbX_3$ unit, and the effect of breaking periodicity by introducing a B-site dopant (blue circle). **b,** Left: Schematic representation of the Bloch states formed at the Γ-point and the R-point to build the valence and the conduction band, respectively. The yellow double arrows highlight the bandwidth of the bands, whereas the grey double arrow shows the direct electronic band gap.

Right: Schematic representation of the same states, now showing the relative change of the electronic energy when a B-site dopant is incorporated. It becomes clear that 1) the band gap widens because of stabilization and destabilization of the VBM and the CBM, respectively, and that 2) the bandwidth for both energy bands is reduced upon doping.